# The effect of ionization on the populations of excited levels of C IV and C V in tokamak edge plasmas

K D Lawson[1], I H Coffey[2], K M Aggarwal[2], F P Keenan[2] & JET-EFDA Contributors*

*JET-EFDA, Culham Science Centre, Abingdon, OX14 3DB, UK*
[1] *Euratom/CCFE Fusion Association, Culham Science Centre, Abingdon, OX14 3DB, UK*
[2] *Astrophysics Research Centre, School of Mathematics and Physics, Queen's University Belfast, Belfast, BT7 1NN, Northern Ireland, UK*

## Abstract

The main populating and depopulating mechanisms of the excited energy levels of ions in plasmas with densities $<10^{23}$-$10^{24}$ m$^{-3}$ are electron collisional excitation from the ion's ground state and radiative decay, respectively, with the majority of the electron population being in the ground state of the ionization stage. Electron collisional ionization is predominately expected to take place from one ground state to that of the next higher ionization stage. However, the question arises as to whether, in some cases, ionization can also affect the excited level populations. This would apply particularly to those cases involving transient events such as impurity influxes in a laboratory plasma. An analysis of the importance of ionization in populating the excited levels of ions in plasmas typical of those found in the edge of tokamaks is undertaken for the C IV and C V ionization stages. The emphasis is on those energy levels giving rise to transitions of most use for diagnostic purposes (n≤5). Carbon is chosen since it is an important contaminant of JET plasmas; it was the dominant low Z impurity before the installation of the ITER-like wall and is still present in the plasma after its installation. Direct electron collisional ionization both from and to excited levels is considered. Distorted-wave Flexible Atomic Code calculations are performed to generate the required ionization cross sections, due to a lack of atomic data in the literature. Employing these data, ionization from excited level populations is not found to be significant in comparison with radiative decay. However, for some energy levels, ionization terminating in the excited level has an effect in the steady-state of the order of the measurement errors (±10%). During transient events, ionization to excited levels will be of more importance and must be taken into account in the calculation of excited level populations. More accurate atomic data, including possible resonance contributions to the cross sections, would tend to increase further the importance of these effects.

## 1. Introduction

In laboratory plasmas for which $n_e < 10^{23}$-$10^{24}$ m$^{-3}$, such as those found in tokamaks, the excited level populations of an ion are determined largely by electron collisions and spontaneous radiative decay. At higher densities, collisions increasingly dominate the populating mechanisms with the result that the level populations begin to approach a statistical distribution. The passive radiation emitted from the low density plasmas provides information on the excited level populations. Through their description in a collisional-radiative model the populations can be used for a variety of diagnostic purposes, such as determining the impurity content of the plasma, the measurement of certain plasma parameters and the study of impurity transport. An understanding of impurity transport is crucial if the impurity behaviour in next-step devices, such as ITER, is to be predicted (Loarte *et al.* 2007).

The chief processes included in the collisional-radiative models used are illustrated in figure 1, which shows a schematic diagram of energy levels in two adjacent ionization stages. Populating and depopulating channels of level 2 in the higher ionization stage are illustrated for a low density plasma, $n_e \sim 10^{19} m^{-3}$, in which the vast majority of the population is in the ground state, g, and the excited level populations are in effect collisionally decoupled. The dominant populating channel is electron collisional excitation from the ground state. If the ionization stage contains metastable levels, from which there are no allowed radiative transitions to the ground state, electron collisions from these levels must also be considered, since they can contain significant populations. Depopulation is by radiative decay; for low Z elements there is emission in the XUV/VUV spectral regions and usually some visible lines. Even in the low density model, radiative cascading from higher levels should be included. Lawson *et al.* (2008), who compared theoretical and modelled carbon line intensity ratios, find that this can lead to a ~10% increase in certain C IV excited level populations.

Two further processes which should be considered are ionization and recombination. The latter must be included for an accurate description of divertor plasmas. For example, Lawson *et al.* (2011) compared C IV theoretical line intensity ratios with a database of steady-state, divertor measurements from JET and found agreement to within the experimental accuracy of ~±10%. These authors showed that charge exchange recombination contributes between 11 and 21% to the C IV $1s^2 3p$ $^2P$ level populations, although for the large database considered free electron recombination appears to be small. Fenstermacher *et al.* (1997) demonstrated the importance of recombination in radiative divertor plasmas in DIII-D produced with either D gas-puffing or N or Ne impurity seeding. Similar D gas-puffing experiments were carried out by Nakano *et al.* (2007, 2009) in JT-60 and they again emphasize the importance of recombination in detached divertor plasmas. However, these last authors do not apply their collisional-radiative model in a self-consistent way, which may lead to inaccuracies in their analysis.

The importance of recombination in the main chamber Scrape-off-Layer (SOL) is less clear, since there is a lack of consistency between theory and measurements. This suggests that the collisional-radiative model used to describe this plasma region is incomplete or that there is some other aspect of the analysis which is not understood (Lawson *et al.* 2012). Certainly, recombination (free electron or charge exchange) does not explain the observed disagreement and its inclusion tends to make the discrepancy worse (Lawson *et al.* 2008). Indeed, ionization might be expected to be the more important process in the main chamber SOL.

The ionization rate coefficients for transitions from one ground state to that of the next higher ionization stage are much larger than to those to excited levels. Consequently, the assumption is generally made that ionization takes place from one ground state to the next and has no significant effect on excited state populations. This is indeed consistent with the steady-state divertor measurements of Lawson *et al.* (2011). Nevertheless, Lawson *et al.* (2012) found evidence that this may not always be the case, particularly during transient events such as an impurity influx, which may even affect C excited level populations within the divertor. The study of transient events is of particular interest in impurity transport modelling.

The radial impurity transport in a tokamak can be parameterized in terms of a diffusion coefficient, $D$, and convective velocity, $v_r$, the radial particle flux for an ionization stage z being

$$\Gamma_z(r,t) = -D(r,t)\frac{dn_z(r,t)}{dr} + v_r(r,t)\, n_z(r,t),$$

where $n_z$ is the ion density for this stage. Under steady-state conditions, when there is no net impurity flux, it is only possible to determine the ratio of $D$ and $v_r$. To allow separate measurements of these parameters, transient events (such as an impurity influx) must be analyzed. In the core plasma this technique is widely employed to study transport, either using laser ablation techniques or gas-puffing to inject impurities (e.g. Giroud *et al.* 2007). During an influx, electron collisional ionization becomes an important mechanism, since the ions are moved to higher temperature regions than would be normal under ionization equilibrium.

Hence, we examine the assumption that ionization has no significant effect on the excited level populations via a more detailed study of the direct electron collisional ionization of the C III to C V ionization stages, to assess its impact on the C IV and C V excited level populations. Previously, carbon was the main low Z impurity in the JET tokamak and residual levels have already been detected in the upgraded machine (Brezinsek *et al.* 2012) with its ITER-like Be and W plasma-facing surfaces (Matthews *et al.* 2011). In the JET tokamak, the C IV ion stage falls in the plasma edge, outside the separatrix, whereas C V will usually occur at higher temperatures straddling the separatrix. Both ionization to and from excited levels are considered in our analysis, allowing its effect to be assessed both during transient and steady-state operations. It is noted that excitation followed by autoionization is not expected to be as important in low Z ions as direct ionization and therefore is not considered in the present analysis (e.g. for C IV see Crandall *et al.* 1979).

To date, the most important use of ionization data has been in determining the ionization balance between different ionization stages. Consequently, the majority of published cross sections and ionization rate coefficients deal with ionization from ground or metastable levels, with little data for ionization either starting or terminating in excited levels. A few examples of the available data for the ionization stages of interest are given in section 2 and the importance of ionization *from* excited C IV and C V levels is discussed in section 3. To assess ionization to excited levels, it is necessary to develop population models and have estimates of the ratio of the ground state populations of the initial and final ionization stages. The population models are described in section 4, where we also present measurements of the ground state population ratios. Section 5 assesses the ionization *to* the C IV and C V excited states, while the results are discussed in section 6. Finally, the conclusions are given in section 7.

## 2. Published ionization data for C III to C V

There are a number of publications dealing with the ionization of the C III to C V ionization stages. However, all but one present total cross sections or rates, or only deal with ionization from or to the ground or metastable levels. Such data are essential in enabling the ionization balance to be determined, but do not allow any assessment of the effect of ionization on excited states. For example, Bell *et al.* (1983) and Dere (2007) provide data for a range of elements and ionization stages, in both cases reviewing the available calculations and measurements. The former give recommended cross sections and ionization rate coefficients for the light elements up to O. Dere's subsequent review includes all elements up to Zn and, where no calculations are available, the Flexible Atomic Code (FAC) of Gu (2003) is used to generate the necessary data. He considers inner and outer shell ionization and excitation-autoionization (EA) to give ionization rate coefficients for all ionization stages of all elements under consideration. The direct ionization (DI) is calculated using the distorted-wave option of FAC. Suno and Kato (2005) review the available data for all ionization stages of carbon making recommendations as to the preferred ionization cross sections. This is part of a study that includes electron collisional excitation and charge exchange recombination in a cross section database for carbon.

Younger (1981) and Fogle *et al*. (2008) deal specifically with Be-like ions, the former carrying out distorted-wave calculations with exchange. Fogle *et al*. make a comparison between new experimental measurements and the available theoretical data for C III, N IV and O V including their own R-matrix with pseudostates calculations. The last are found to give the best agreement with experiment. Both sets of authors consider ionization from the $1s^2 2s^2$ $^1S_0$ ground state and from the $1s^2 2s2p$ $^3P$ levels, the latter being metastable since radiative decays to the ground state are spin forbidden.

Ionization cross sections for H- and He-like ions are provided by Fang *et al*. (1995). They use distorted-wave with exchange calculations, which include a relativistic correction. Mitnik *et al*. (1999) deal only with Li-like C and calculate ionization cross sections using the R-matrix with pseudostates method. The accuracy of the calculation when a small pseudostate basis is used is investigated by Badnell and Griffin (2000) for Li-like ions including C IV. Sampson and Zhang (1988) derive rate coefficients for innershell ionization of a number of Li-like ions, leading to population in the first excited configuration of the relevant He-like ions. However, this configuration only contains two levels, $1s2s$ $^3S_1$ and $^1S_0$, which are again metastable with no allowed radiative transition to the $1s^2$ $^1S_0$ ground state.

The emphasis of the above papers is on the direct ionization (DI) process, which provides the main contribution to the total ionization cross section for the ions of interest. Two publications that deal specifically with indirect ionization processes are those of Scott *et al*. (2000) and Knopp *et al*. (2001). The three indirect processes that need to be considered for C IV are excitation-autoionization (EA), resonant excitation double autoionization (REDA) and resonant excitation auto-double-ionization (READI). In EA, an innershell excitation is followed by autoionization. Both REDA and READI involve the resonant capture of the incident electron and the subsequent emission of two electrons. In the REDA process the electrons are sequentially emitted, whereas in READI the emission is simultaneous. Resonances due to the REDA process occur at energies above the EA threshold, whereas those due to READI can occur at energies below. For the ions of interest in the present study, these processes are expected to be small (although significant), while for higher Z elements (e.g. Fe XVI) they can dominate the ionization cross section (Linkeman *et al*. 1995). Scott *et al*. use the R-matrix with pseudostates method to investigate the EA and REDA processes as well as the DI. All three indirect processes are treated by Knopp *et al*. using the unified R-matrix approach of Berrington *et al*. (1997), which is again an R-matrix with pseudostates method. Interference between the interacting EA and REDA channels and between the READI and DI channels, which are experimentally observed, are satisfactorily explained by the theory. Unfortunately, neither of these papers give an indication of the terminating states of the ionization processes.

Pindzola *et al*. (2011) present the only results for the ions of interest for ionization involving excited states that are not metastable. Nonperturbative R-matrix and perturbative distorted-wave calculations of the direct electron impact ionization from the C IV $1s^2 5s$ level are compared and found to be in reasonable agreement for this moderately charged ion. Distorted-wave calculations are also presented for ionization from the bundled $1s^2 5l$ configurations, where $l$ = 0 - 4.

## 3. Ionization *from* excited C IV and C V levels

Radiative decay is the main depopulating mechanism in the low density edge plasmas of JET, for which $n_e \sim 10^{19}$ m$^{-3}$. Therefore, in assessing the importance of electron collisional ionization from excited C IV and C V levels, it is necessary to compare the depletion of population due to electron collisional ionization, $n_e n_i s_{ik}$, from level $i$ to a level $k$ in C V or C VI with that due to the sum of radiative decays, $\Sigma n_i A_{ij}$, from level $i$ to lower levels denoted by $j$ within the same ion, where $s_{ik}$ and $A_{ij}$ are the ionization rate coefficient and radiative transition probability, respectively.

Given the lack of direct electron collisional ionization rate coefficients in the literature, the required data were generated using the FAC code of Gu (2003). In the calculations to assess ionization from excited C IV levels, all C IV configurations up to and including the $n$ = 5 shell were included, the $n$ = 1 shell being taken to be closed. The populations in levels with an incomplete $n$ = 1 shell are very small in tokamaks, with transitions from these levels not normally being observed. It is also necessary to define the C V configurations in which the ionizations terminate. For this stage, the ground configuration, $1s^2$, and excited configurations $mlrl'$, where $m$ = 1, 2 and $r$ = 2, 3, 4, 5, were included. Since the calculations are to be used to demonstrate if electron collisional ionization is important, the distorted-wave option within FAC was used. This gives cross sections which tend to be higher than the other FAC ionization calculations and therefore will provide the most stringent test. Ionization cross sections were taken from the FAC output and rate coefficients were derived. The ionization rate coefficient from state $j$ to state $k$ is

$$s_{jk} = \frac{2.550 \times 10^{-14}}{\omega_j T_e^{3/2}} \int_{E_{th}}^{\infty} \exp\left(\frac{-E_i}{T_e}\right) \Omega_{jk} \, dE_i \ \ m^3 s^{-1},$$

where $\omega_j$ is the statistical weight of state $j$, $E_i$ and $E_{th}$ are the incident electron and threshold energies and $\Omega_{jk}$ is the ionization collision strength given by

$$\Omega_{jk} = \frac{Q_{jk}}{a_o^2} \omega_j \frac{E_i}{I_H}.$$

Here $Q_{jk}$ is the ionization cross section, $a_o$ the Bohr radius and $I_H$ the ionization potential of H. The electron temperature $T_e$ and $E_i$ are measured in eV.

As expected, the majority of C IV ionization is to the C V ground state, accounting for over 99.7% of the total rate at the temperatures of interest, 15 to 70 eV. Since the ionization rate is largest at high temperatures, table 1 compares the sum of the A-values from the lowest 14 excited C IV levels with the total ionization rates for the worst case (at 70 eV) and at an electron density of $10^{19}$ m$^{-3}$. The A-values used in this table are from Aggarwal and Keenan (2004). It can be seen that the direct electron collisional ionization from the C IV excited levels at this density is small, at most being ~$5\times10^{-5}$ of the loss of excited state population through radiative decay.

Table 1. Comparison of radiative decay and total ionization rates from C IV levels.

| Energy level | $\Sigma A$-values (s$^{-1}$) | $\Sigma n_e s$ at $n_e$ of $10^{19}$ m$^{-3}$ (s$^{-1}$) | Fractional depletion due to ionization |
|---|---|---|---|
| $1s^2 2p\ ^2P_{1/2}$ | 2.76e+8 | 1.36e+4 | 4.93e-5 |
| $1s^2 2p\ ^2P_{3/2}$ | 2.78e+8 | 1.34e+4 | 4.82e-5 |
| $1s^2 3s\ ^2S_{1/2}$ | 4.22e+9 | 3.90e+4 | 9.24e-6 |
| $1s^2 3p\ ^2P_{1/2}$ | 4.55e+9 | 5.17e+4 | 1.14e-5 |
| $1s^2 3p\ ^2P_{3/2}$ | 4.54e+9 | 5.17e+4 | 1.14e-5 |
| $1s^2 3d\ ^2D_{3/2}$ | 1.75e+10 | 7.63e+4 | 4.37e-6 |
| $1s^2 3d\ ^2D_{5/2}$ | 1.75e+10 | 7.52e+4 | 4.31e-6 |
| $1s^2 4s\ ^2S_{1/2}$ | 2.64e+9 | 8.97e+4 | 3.40e-5 |
| $1s^2 4p\ ^2P_{1/2}$ | 2.85e+9 | 7.93e+4 | 2.78e-5 |
| $1s^2 4p\ ^2P_{3/2}$ | 2.85e+9 | 7.95e+4 | 2.79e-5 |
| $1s^2 4d\ ^2D_{3/2}$ | 7.61e+9 | 1.40e+5 | 1.84e-5 |
| $1s^2 4d\ ^2D_{5/2}$ | 7.61e+9 | 1.56e+5 | 2.05e-5 |
| $1s^2 4f\ ^2F_{5/2}$ | 3.54e+9 | 1.90e+5 | 5.38e-5 |
| $1s^2 4f\ ^2F_{7/2}$ | 3.54e+9 | 1.90e+5 | 5.38e-5 |

A similar analysis was undertaken for electron collisional ionization from the C V excited levels. In this case, no published data are available and so calculations were performed using the distorted-wave option of FAC. The ground configuration for C V, $1s^2$, and excited configurations $mlrl'$, where $m = 1, 2$ and $r = 2, 3, 4, 5$, were included in the calculation, while for the final C VI levels all configurations up to and including $n = 5$ were used. Ionization rates at an electron density of $10^{19}$ m$^{-3}$ were derived from the output ionization cross sections and comparisons made with the sums of the transition probabilities taken from Aggarwal *et al.* (2011). Table 2 compares the A-value and ionization rate sums of the levels for which ionization has the greatest effect amongst the lowest 30 levels of C V. Again the ionization rate is calculated at a high temperature, 400 eV, this corresponding to the worst case (i.e. largest ionization rate) expected for C V.

It may be seen from table 2 that the fractional depletions can be somewhat higher than those for C IV, but are still so small that electron collisional ionization from the excited levels does not play a significant role in the C V excited state population balance. However, unlike C IV, C V has two metastable levels, $1s2s\ ^3S_1$ and $^1S_0$, and as might be expected these are exceptions. There are no allowed radiative transitions to lower levels, the transition probabilities being very small. Depopulation will occur through electron and heavy particle collisional de-excitation or excitation rather than through radiative decay. The rates for these processes are closer to the ionization rates. For example, the largest electron collisional rates from the $1s2s\ ^3S_1$ and $1s2s\ ^1S_0$ metastable levels are both due to excitation rather than de-excitation. These are for the $1s2s\ ^3S_1$ - $1s2p\ ^3P_2$ and $1s2s\ ^1S_0$ - $1s2p\ ^1P_1$ transitions, with values of $2.66\times10^5$ and $5.77\times10^5$ s$^{-1}$, respectively, at a density of $10^{19}$ m$^{-3}$. The excitation rates are calculated from the R-matrix data of Aggarwal *et al.* (2011) at the highest temperature (86 eV) for which their results are given. The corresponding ionization rates from these levels at this temperature are $4.88\times10^3$ and $5.56\times10^3$ s$^{-1}$, i.e. ~1% of the excitation rates. In this case, the rates are sufficiently close to warrant further calculations and consequently the two additional options within FAC have been checked. The first is a Coulomb-Born approximation, in which radial integrals are obtained from tables of the Coulomb-Born-exchange results of Golden and Sampson (1977, 1980). In the second option, there is recognition that the distorted-wave approximation tends to overestimate the cross sections near threshold and hence the binary-encounter-dipole (BED) theory of Kim and Rudd (1994) is used. The total ionization cross sections in this approach are scaled by a factor $E_i/(E_i+E_{th})$, where $E_i$ is the energy of the incident electron and $E_{th}$ is the ionization threshold energy, thereby reducing the near-threshold cross sections. These calculations yield rates within 50% of the distorted-wave results, which, in this case, are intermediate to the other results. For the Coulomb-Born calculation, ionization rates of $6.65\times10^3$ and $8.30\times10^3$ s$^{-1}$ from the 1s2s $^3S_1$ and 1s2s $^1S_0$ levels, respectively, are found at a density of $10^{19}$ m$^{-3}$ and a temperature of 86 eV, whereas the BED method gives values of $4.24\times10^3$ and $3.87\times10^3$ s$^{-1}$, respectively.

Table 2. Comparison of radiative decay and total ionization rates from C V levels.

| Energy level | $\Sigma A$-values (s$^{-1}$) | $\Sigma n_e s$ at $n_e$ of $10^{19}$ m$^{-3}$ (s$^{-1}$) | Fractional depletion due to ionization |
|---|---|---|---|
| 1s2s $^3S_1$ | 4.19e+1 | 1.16e+4 | Metastable |
| 1s2s $^1S_0$ | 1.75e+0 | 1.20e+4 | Metastable |
| 1s2p $^3P_1$ | 8.79e+7 | 1.38e+4 | 1.57e-4 |
| 1s2p $^3P_0$ | 6.02e+7 | 1.39e+4 | 2.30e-4 |
| 1s2p $^3P_2$ | 6.08e+7 | 1.37e+4 | 2.26e-4 |
| 1s4s $^3S_1$ | 3.90e+9 | 4.52e+4 | 1.16e-5 |
| 1s4s $^1S_0$ | 5.21e+9 | 3.30e+4 | 6.34e-6 |
| 1s4p $^3P_1$ | 7.60e+9 | 4.99e+4 | 6.56e-6 |
| 1s4p $^3P_0$ | 7.60e+9 | 4.99e+4 | 6.57e-6 |
| 1s4p $^3P_2$ | 7.59e+9 | 5.00e+4 | 6.58e-6 |
| 1s4f $^1F_3$ | 8.62e+9 | 1.02e+5 | 1.19e-5 |
| 1s4f $^3F_3$ | 8.63e+9 | 1.03e+5 | 1.20e-5 |
| 1s4f $^3F_4$ | 8.63e+9 | 1.02e+5 | 1.18e-5 |
| 1s4f $^3F_2$ | 8.63e+9 | 1.03e+5 | 1.20e-5 |

It is noted that, as with the ionization of C IV, a large proportion of ionizations terminate in the ground state of the next higher ionization stage, the exact proportion depending on the C V configuration. From the 1s2$l$ configurations ~95% of ionization ends in the C VI ground state, whereas the proportion is even higher, ~98.6% and 99.6%, from the 1s3$l$ and 1s4$l$ configurations, respectively.

## 4. Population models

### a. C III

To assess the importance of ionization to excited levels, it is necessary to develop a model for the energy level populations of both the initial and final ionization stages. At the low densities typical of tokamak plasmas, the higher-lying energy levels have very small populations. Collisions from these levels therefore contribute little to the population of adjacent energy levels or, of relevance in the present context, to ionization. Consequently, accurate modelling of the populations of the lower energy levels is of most importance. Nevertheless, a sufficient number of energy levels should be included in the calculation to give an indication of the relative importance of the different shells towards ionization. This also ensures that radiative cascading from higher levels, which increases somewhat the lower level populations, can be properly treated. For example, in the C IV population calculations, increases of ~10% due to radiative cascading are typical (Lawson *et al*. 2008).

A collisional-radiative calculation was performed for the lowest 20 fine-structure C III energy levels, which included electron collisional excitation and de-excitation among all levels and all significant radiative decays. For an excited level *i*, the population, $n_i$, is determined from the rate equations

$$\frac{dn_i}{dt} = n_e \sum_{j \neq i} n_j q_{ji} - n_e \sum_{j \neq i} n_i q_{ij} + \sum_{j>i} n_j A_{ji} - \sum_{j<i} n_i A_{ij},$$

where $A_{ij}$ is the Einstein spontaneous emission transition probability and $q_{ij}$ the collisional excitation or de-excitation rate from level *i* to *j*. Note that in this representation of the collisional-radiative model source terms such as ionization and recombination are omitted. In the steady-state approximation,

$$\frac{dn_i}{dt} = 0,$$

leading to a set of simultaneous equations, which can be solved for $n_i / n_g$, where $n_g$ is the ground level population. Hence, in addition to the transition probabilities, collisional excitation and de-excitation rates are required. For ions such as C III to C V, the main populating mechanism is electron collisions.

The lowest 20 C III levels include the 10 within the *n* = 2 shell plus the 2s3*l* levels, where *l* = s, p and d. An additional 30 higher-lying levels expected to have the most significant populations were also included in the model. These fall within the 2*lml'* configurations, where *l* = s and p, *m* = 4 and 5 and *l'* = s, p, d, and f. For the additional levels all significant radiative decays were included, but only excitation from low levels that have significant populations. Normally this would be the ground state, in this case $1s^2 2s^2$ $^1S_0$. However, C III also has metastable levels that can develop significant populations; these are within the $1s^2 2s2p$ $^3P$ multiplet, from which radiative decays to the ground state are spin forbidden. The A-values are therefore small, that for the $1s^2 2s^2$ $^1S_0$ - $1s^2 2s2p$ $^3P_1$ being 114 s$^{-1}$ with all other radiative rates from the metastable levels being orders of magnitude smaller. Depopulation of the metastable levels is achieved by collisions rather than radiative decay. Both electron and heavy particle collisional excitation and de-excitation must be considered and D fuel excitation and de-excitation rates were included for transitions between the metastable levels in order to represent the decay channels from these levels as accurately as possible. For other levels, radiative decay is the dominating depopulating mechanism at the densities considered and so electron collisional de-excitation of the 30 higher-lying levels was largely neglected. The model did include collisional de-excitation to the ground and metastable levels, but these channels were not found to be significant.

Radiative transition probabilities for C III were taken from the NIST compilation (Ralchenko *et al*. 2011), while heavy particle collisional excitation rates for transitions within the $1s^2 2s2p$ $^3P$ multiplet for a number of Be-like ions have been calculated by Ryans *et al*.

(1998). Two R-matrix calculations have been performed for electron excitation collision strengths for C III. The first, reported by Berrington (1985) and Berrington *et al*. (1989), involves a 12 state calculation corresponding to the lowest 20 fine-structure levels. Fine-structure effective collision strengths have been derived by Lang as described in the C III collisional excitation ADAS files (Summers 2004). The ADAS files also contain effective collision strengths for a number of additional LS-resolved levels, although these data are expected to be less accurate than the R-matrix results. We obtained j-resolved effective collision strengths for these higher additional levels using the corresponding splittings to those provided by Lang for the ADAS N IV collisional excitation files.

The second R-matrix calculation is a 238 term one with pseudostates by Mitnik *et al*. (2003). It includes 90 terms from both within the *n* = 2 shell and those within the 2*lml'* configurations, where *m* = 3 to 5. The fine-structure effective collision strengths were obtained from the LS-resolved data using the same splittings as for the Berrington dataset. This allowed a j-resolved calculation to be undertaken and hence the most accurate description of the metastable levels. The population calculations were performed at an electron density of $10^{19}$ m$^{-3}$ and a D ion density of $8\times10^{18}$ m$^{-3}$. Both datasets have been used, although the later and more complete set of Mitnik *et al.* are preferred. For many of the levels there is agreement between the two calculations to better than ~15%, although in some cases, for example the 2p3p $^3$D and 2s4d $^3$D levels, the populations differ by more than a factor of 2.

Since source terms have been omitted from the collisional-radiative model used, it is of interest to test the preferred population model against experimental measurements, in particular to check the metastable level populations, which can be strongly influenced by these terms. Theoretical line intensity ratios derived from the model have been compared with the emission from the JET divertor. The best agreement between theory and measurement is obtained for metastable level populations that are within ~10% of those predicted. It is noted that, although the spectrometer sensitivity calibration at short wavelengths (λ < 450 Å) was determined independently, the long wavelength sensitivity calibration was varied as part of the optimization procedure used to match the theoretical and measured line intensity ratios.

### b. C IV

In the study presented here a low density model was used for C IV, in which the levels were populated by electron collisional excitation from the ground state and radiative cascading from higher levels and depopulated by radiative decay. When compared with a collisional-radiative population model in which electron collisions among all levels are included, the low density model was found to be accurate to ~1%. R-matrix calculations of electron collisional excitation rates with the DARC code by Aggarwal and Keenan (2004) were used, these authors also giving radiative transition probabilities generated by the GRASP code for all levels up to and including those in the *n* = 5 shell. Collisional atomic data are only available at temperatures up to 129 eV, but since the effective collision strengths are slowly varying these have been extrapolated to ~170 eV. At such high values of temperature C IV is expected to be almost entirely ionized. Level populations were calculated at a density of $10^{19}$ m$^{-3}$.

The model for C IV was well-tested against JET data (Lawson *et al*. 2011). An analysis of emission line intensity ratios derived using this population model showed agreement with values measured from JET divertor spectra to within the experimental accuracy of ~±10%. In this case, an independent spectrometer sensitivity calibration was available throughout the spectral range of the measurements (Lawson *et al.* 2009); it was found to be in excellent agreement with that inferred from the comparison of the C IV observations with theory. When comparing with JET divertor measurements, it was necessary to include charge exchange recombination in the population model to obtain good agreement between theory and experiment. The charge exchange atomic data contained in the ADAS database (Summers, 2004), compiled by Maggi (1996), were employed. This only had a significant effect on the 312.4 Å line, the charge exchange recombination contribution to the upper levels ($1s^23p$ $^3P$) of this transition varying between 11 - 21%. The only other line affected by charge exchange was 384.1 Å, although with a smaller contribution of 0.5 - 4%. It was recognized that the conditions that favour recombination most likely will not be appropriate for an ionizing plasma. Hence, the population model without charge exchange recombination was preferred in the present context. Nevertheless, checks were made to investigate if the inclusion of charge exchange recombination made a difference to the conclusions. To account for recombination the populations of the $1s^23p$ $^3P$ levels were increased by 12%. The analysis of the JET divertor emission showed no dependence on free electron recombination and hence it has not been taken into account.

**c. C V**

Although the effect of the ionization of C V on C VI excited levels is not considered in the present study, a population model for C V is nevertheless required to derive excitation Photon Emission Coefficients (PECs) for the C V ionization stage. As will be explained in section 5, this provides a convenient way of assessing the importance of the contribution of the C IV ionization to the C V excited levels.

It was noted in section 3 that C V has two metastable levels, $1s2s$ $^3S_1$ and $^1S_0$, and so as to describe the decay channels from these states as accurately as possible collisional excitation and de-excitation was included for all transitions among the lowest 8 levels. For higher levels only electron collisional excitation from (and de-excitation to) the lowest 3 levels (the ground and the two metastable ones) has been included in the model. These levels have the highest populations and therefore provide the dominant populating channels. A second populating mechanism that must be included is radiative cascading from higher levels; as already explained this can provide a small, but significant, contribution to the populations.

Atomic data for both radiative transition probabilities and R-matrix electron collision strengths were taken from Aggarwal et al. (2011), with the level energies used being those from the NIST compilation (Ralchenko *et al*. 2011). As in the C III population model, heavy particle collisions between close-lying levels having significant populations must be considered. However, the necessary atomic data is not available for the C V ionization stage. Instead, their importance was assessed by artificially increasing the electron collisional rates, in particular to see if heavy particle collisions could deplete the metastable $1s2s$ $^1S_0$ level population, this level lying close to the $1s2s$ $^3P$ multiplet. Increasing the electron rates by a factor of 3 for transitions between the $1s2s$ $^1S_0$ level and the neighbouring $1s2p$ $^3P$ multiplet and between levels within the multiplet itself had only a small effect on the derived populations. For example, that for the $1s2s$ $^1S_0$ level was decreased by between 3 and 4% depending on the temperature, with other changes being ~1% or less. The full details of both the C III and C V population models and their comparison with experimental measurements will be described in future publications.

### d. Ratio of ground state populations

As will be explained in section 5, a further parameter required to assess the importance of ionization to excited states is the ratio of the ground state populations of the initial and final ionization stages, $n_{g-1} / n_g$. The steady-state case is considered first and then changes in the ratio during C impurity influxes are illustrated. In steady-state, the $n_{g-1} / n_g$ ratio for the C III and C IV ionization stages is expected to be ~1, since the differences between the C II, C III and C IV ionization potentials (24, 48 and 64 eV, respectively) are similar. This is confirmed for JET pulse 69931 using the C III and C IV lines at 386.2 and 384.1 Å, respectively. These lines are measured with the JET divertor-viewing double SPRED survey spectrometer (Wolf *et al*. 1995) and the $n_{g-1} / n_g$ ratio determined using the population models described in sections 4a and 4b. It is noted that this calculation can only be made for the divertor plasma, since there is uncertainty regarding the description of the impurity radiation in the main chamber SOL. In contrast, the C V ionization stage has a wide spatial extent due to the high potential (392 eV) required to ionize the $1s^2$ closed shell. Hence, the $n_{g-1} / n_g$ ratio determined for line integrated measurements is expected to be less than 1. Using the C IV and C V lines at 244.9 and 227.2 Å, respectively, and the population models described in sections 4b and 4c gives an estimate for the line-integrated ratio of ~0.2 - 0.5 for pulse 69931. Again, the analysis is carried out for the divertor plasma. If spatially resolved measurements are possible, then $n_{g-1} / n_g$ will be ~1 in the lower temperature, outer regions of the C V emission shell.

The change in the $n_{g-1} / n_g$ ratio during an impurity influx varies widely from influx to influx, in particular depending on the location of the source relative to the observing spectrometer. Consequently, several influxes have been studied to illustrate the range of values that might be expected. Ideally, transitions whose upper levels are strongly coupled to the ground states are used, such as those at 977.0, 312.4 and 40.3 Å for the C III, C IV and C V ionization stages, respectively. However, there are a number of pulses when no measurement of the C V 40.3 Å line is possible and instead C V at 227.2 Å is used. In this case, excitation from the $1s2s$ $^3S_1$ and $^1S_0$ metastable levels should be considered. However, the population calculation of section 4c shows that, at the temperatures of interest, the C V metastable level populations are small. Hence, the dominant populating channel is still electron excitation from the C V ground state, $1s^2$ $^1S_0$, and this allows an estimate of the ground state population ratio. The change in the line intensity ratios immediately before and after a sudden impurity influx, when disturbances to the background plasma are minimal, is taken to represent the change in the ground state population ratio.

In table 3 we present measurements made with the double SPRED spectrometer (Wolf *et al*. 1995) and with the JET single SPRED (Fonck *et al*. 1982) and Schwob-Fraenkel (Schwob *et al*. 1987) survey spectrometers. The double SPRED has a vertical line-of-sight, which for these pulses is aligned to view wall tiles just outside the divertor throat, while the other two instruments have near-midplane horizontal lines-of-sight. All vertical C V measurements relied on the 227.2 Å line, which is clearly resolved and usually unblended in the double SPRED spectrum. The horizontal C V measurements in pulses ≥ 75060 also used this line. However, the spectral resolution of the single SPRED is poorer, making the measurement of the 227.2 Å transition more difficult and less reliable. No evidence of its intensity varying by more than a factor of 2 was found and, hence, a range of values corresponding to the 227.2 Å line intensity increasing by a factor of between 1 and 2 is given. In two pulses, 75478 and 75481, the horizontal C V 227.2 Å feature is blended with intense metal lines and no measurement is possible. The estimates given in table 3 suggest that although small increases $\leq \times 5$ in the $n_{g-1} / n_g$ ratio are most frequently observed, larger increases of up to $\sim \times 10$ are possible. Also listed in the table is the magnitude of the influx, indicated by the increase in the C III or, if not available, the C IV line intensity. This shows that there is only a weak correlation between the size of the influx and the change in the ground state population ratio.

Table 3. Change in the ratio of the ground state populations, $n_{g-1} / n_g$, during various impurity influxes, together with the increase in the C III (or C IV) intensity indicating the magnitude of the influx.

| Pulse | Time (s) | Line-of-sight | | | | | |
|---|---|---|---|---|---|---|---|
| | | Horizontal | | | Vertical | | |
| | | Increase in C III intensity | Change in $n_{g-1} / n_g$ | | Increase in C III (C IV) intensity | Change in $n_{g-1} / n_g$ | |
| | | | C III / C IV | C IV / C V | | C III / C IV | C IV / C V |
| 69510 | 57.1 | 4.4 | 5.8 | 2.9 | (1.4) | - | 2.4 |
| 69522 | 60.6 | 3.0 | 0.8 | 1.3 | (6.3) | - | 1.9 |
| 69522 | 68.6 | 3.7 | 2.6 | 1.0 | (0.4) | - | 1.0 |
| 69849 | 64.7 | 1.9 | 1.2 | 0.8 | (11.0) | - | 6.9 |
| 69849 | 73.4 | 9.6 | 3.4 | 2.2 | - | - | - |
| 69856 | 73.4 | 6.4 | 1.9 | 2.6 | - | - | - |
| 75060 | 69.0 | 3.0 | 1.1 | 1.4-2.8 | 3.9 | 2.1 | 1.7 |
| 75478 | 58.4 | 3.7 | 1.1 | C V blend | 14.0 | 1.3 | 5.0 |
| 75481 | 59.2 | 2.4 | 0.7 | C V blend | 7.4 | 1.6 | 1.4 |
| 79721 | 64.7 | 14.0 | 1.2 | 6.0-12.0 | - | - | - |
| 79726 | 66.0 | 17.0 | 4.6 | 1.8-3.7 | - | - | - |
| 79727 | 66.0 | 20.0 | 3.7 | 2.6-5.2 | - | - | - |
| 79747 | 66.2 | 13.0 | 1.4 | 4.5-9.0 | - | - | - |
| 79756 | 68.7 | 10.0 | 7.5 | 1.0-1.4 | - | - | - |

# 5. Ionization *to* excited C IV and C V levels

## a. C IV

As before, the required ionization rate coefficients were generated using FAC, the three calculations being described in section 3. In the following tables, the results of the distorted-wave calculations are used, although comparisons were made with the other two. The ionization cross sections output by FAC were converted to ionization rates as described in section 3.

In the calculations, the $n = 1$ shell was taken to be closed. For the final C IV ion stage all levels within the $n = 2$ to 5 shells were included. Those considered for the initial C III stage were all levels in the $2lml'$ configurations, where $l = $ s and p, $m = 2 - 5$ and $l'$ can take values s, p, d, f and g depending on the value of $m$. In total, 166 C III levels were treated in the calculation, among which were the 50 levels of the population calculation. An electron density of $10^{19}$ m$^{-3}$ and a D ion density of $8 \times 10^{18}$ m$^{-3}$ were employed in the population calculations, which used the electron collisional excitation data of Mitnik *et al*. (2003).

Table 4. Ionization contributions to the individual C IV levels, $\chi_{IV}$, and their fractions of the total ionization contributions, $\chi_{tot}$, together with populations of the C IV levels as a fraction of the ground level population, $n_g$, at $T_e = 20$ and 70 eV.

| C IV Level index | C IV Level | 20eV $\chi_{tot} = 2.22\text{e-}15\ \text{m}^3\text{s}^{-1}$ | | | 70eV $\chi_{tot} = 1.48\text{e-}14\ \text{m}^3\text{s}^{-1}$ | | |
|---|---|---|---|---|---|---|---|
| | | $\chi_{IV}$ (m$^3$s$^{-1}$) | $\chi_{IV}/\chi_{tot}$ | $n_i/n_g$ | $\chi_{IV}$ (m$^3$s$^{-1}$) | $\chi_{IV}/\chi_{tot}$ | $n_i/n_g$ |
| 1 | $1s^22s\ ^2S_{1/2}$ | 1.72e-15 | 7.8e-1 | 1.00 | 1.05e-14 | 7.1e-1 | 1.00 |
| 2 | $1s^22p\ ^2P_{1/2}$ | 1.64e-16 | 7.4e-2 | 8.02e-4 | 1.42e-15 | 9.6e-2 | 7.53e-4 |
| 3 | $1s^22p\ ^2P_{3/2}$ | 3.28e-16 | 1.5e-1 | 1.59e-3 | 2.83e-15 | 1.9e-1 | 1.50e-3 |
| 4 | $1s^23s\ ^2S_{1/2}$ | 9.22e-19 | 4.2e-4 | 1.32e-6 | 2.52e-17 | 1.7e-3 | 2.87e-6 |
| 5 | $1s^23p\ ^2P_{1/2}$ | 2.53e-19 | 1.1e-4 | 2.68e-7 | 7.01e-18 | 4.7e-4 | 7.91e-7 |
| 6 | $1s^23p\ ^2P_{3/2}$ | 5.03e-19 | 2.3e-4 | 5.40e-7 | 1.40e-17 | 9.4e-4 | 1.58e-6 |
| 7 | $1s^23d\ ^2D_{3/2}$ | 8.16e-20 | 3.7e-5 | 1.72e-7 | 2.32e-18 | 1.6e-4 | 5.02e-7 |
| 8 | $1s^23d\ ^2D_{5/2}$ | 1.23e-19 | 5.6e-5 | 2.60e-7 | 3.49e-18 | 2.4e-4 | 7.53e-7 |
| 9 | $1s^24s\ ^2S_{1/2}$ | 4.53e-20 | 2.0e-5 | 2.45e-7 | 1.97e-18 | 1.3e-4 | 7.91e-7 |
| 10 | $1s^24p\ ^2P_{1/2}$ | 1.54e-20 | 6.9e-6 | 3.39e-8 | 6.43e-19 | 4.3e-5 | 1.22e-7 |
| 11 | $1s^24p\ ^2P_{3/2}$ | 3.07e-20 | 1.4e-5 | 1.81e-7 | 1.28e-18 | 8.6e-5 | 6.49e-7 |
| 12 | $1s^24d\ ^2D_{3/2}$ | 5.70e-21 | 2.6e-6 | 5.88e-8 | 2.40e-19 | 1.6e-5 | 2.18e-7 |
| 13 | $1s^24d\ ^2D_{5/2}$ | 8.62e-21 | 3.9e-6 | 8.83e-8 | 3.62e-19 | 2.4e-5 | 3.27e-7 |
| 14 | $1s^24f\ ^2F_{5/2}$ | 3.73e-22 | 1.7e-7 | 7.11e-8 | 1.03e-20 | 7.0e-7 | 1.99e-7 |
| 15 | $1s^24f\ ^2F_{7/2}$ | 4.92e-22 | 2.2e-7 | 9.51e-8 | 1.36e-20 | 9.2e-7 | 2.65e-7 |

Once the C III ionization rate coefficients and the level populations were calculated, the effect of direct collisional ionization on the C IV excited levels could be investigated. Table 4 lists the total ionization contributions from all the C III levels, $k$, to all C IV levels, $i$,

$$\chi_{tot} = \sum_k \sum_i \frac{n_k}{n_{g-1}} s_{ki},$$

where $n_{g-1}$ is the ground level population of the initial ionization stage, in this case C III. Also given in the table are the ionization contributions to individual C IV levels,

$$\chi_{IV} = \sum_k \frac{n_k}{n_{g-1}} s_{ki},$$

at temperatures of 20 and 70 eV, along with values of the fraction $\chi_{IV}/\chi_{tot}$. The temperature of 20 eV corresponds to that at which both C III and C IV ions are expected to exist in the plasma under an equilibrium ionization balance, while 70 eV to a more extreme case of the highest temperature at which significant numbers of C IV ions will be found. In practice, the latter will be most important during impurity influxes, when steep ion density gradients push ions to higher temperature plasma regions than would be usual under equilibrium conditions.

It can be seen from table 4 that at both temperatures there is a marked departure from the expectation that the final state following direct ionization is the ground level of C IV and it should be noted that the C IV ionization stage does not contain any metastable levels. Only ~70% of the ionization results in the C IV $1s^22s\ ^2S_{1/2}$ ground state and a significant proportion populates excited levels. For example, ~10% and 20% of the ionization populates the C IV $1s^22p\ ^2P_{1/2}$ and $^2P_{3/2}$ levels, respectively. Even the $1s^23s\ ^2S_{1/2}$, $1s^23p\ ^2P_{1/2}$ and $^2P_{3/2}$ levels receive, respectively, ~0.1%, ~0.03% and ~0.05% of the ionization. Although the latter contributions to the excited C IV populations are small, they have to be viewed in the context of the small populations resulting from electron collisional excitation. The C IV level populations are also given in table 4 and it can be seen that at the electron densities typical of the plasma edge, ~$10^{19}$ m$^{-3}$, the C IV 1s$^2$2p levels have a population of ~$10^{-3}n_g$ and the higher levels $\leq 3\times 10^{-6} n_g$, where $n_g$ is the ground state population. The significant proportion of ionization resulting in excited C IV levels is due not only to the ionization rates, but is enhanced by the C III 1s$^2$2s2p $^3$P metastable levels having significant populations. Typically these are ~$0.2 n_g$, $0.55 n_g$ and $0.9 n_g$ for the 1s$^2$2s2p $^3$P$_0$, $^3$P$_1$ and $^3$P$_2$ levels, respectively, at the temperatures of interest. The importance of the metastable levels is illustrated by table 5, which gives the ionization contributions from individual C III levels to all the C IV levels included in the calculation,

$$\chi_{III} = \sum_i \frac{n_k}{n_{g-1}} s_{ki}.$$

For clarity, $\chi_{III}$ is also given as a fraction of the total ionization contribution in table 5. Since ionization to excited levels is of particular interest the contribution from the individual C III levels to the excited C IV levels alone,

$$\chi_{III}^{ex} = \sum_{i>1} \frac{n_k}{n_{g-1}} s_{ki},$$

is also listed, together with their fractions of the total ionization contributions to the excited C IV levels,

$$\chi_{tot}^{ex} = \sum_k \sum_{i>1} \frac{n_k}{n_{g-1}} s_{ki}.$$

The importance of the metastable levels is clear in that ionization from the 1s$^2$2s2p $^3$P$_2$ level (41%) exceeds that from the ground state 1s$^2$2s$^2$ $^1$S$_0$ (26%). If ionization to excited C IV states is considered then that from the metastable levels 1s$^2$2s2p $^3$P$_{0,1,2}$ is even more important, contributing 10%, 31% and 51%, respectively, compared with only 8% from the ground state.

Although the ionization of C III differs markedly from that expected in a 'ground/metastable state to ground state' scenario, it is still necessary to assess if ionization to the excited C IV states is sufficiently large to influence the populations of those levels. A convenient assessment method is to combine the ionization rates and populations to give an ionization Photon Emission Coefficient (PEC). This follows the procedure used in ADAS for charge exchange and free electron recombination PECs. The ionization PEC for a C IV transition $i$ to $j$ resulting from the ionization of a C III level $k$ is defined as

$$\varepsilon_{ij}^{ion} = \frac{A_{ij}}{\sum_{j1<i} A_{ij1}} \sum_k \frac{n_k}{n_{g-1}} s_{ki},$$

with the resultant line intensity for the transition being

$$I_{ij} = n_e n_g \varepsilon_{ij}^{exc} + n_e n_{g-1} \varepsilon_{ij}^{ion}. \tag{1}$$

In this equation, $\varepsilon_{ij}^{exc}$ is the excitation PEC for transition $i$ to $j$, where

$$\varepsilon_{ij}^{exc} = \frac{n_i A_{ij}}{n_e n_g}.$$

Table 5. Ionization contributions from the individual C III levels to all C IV levels, $\chi_{III}$, and to the excited C IV levels, $\chi^{ex}_{III}$, and their fractions of the total ionization contributions at $T_e$ = 20 and 70 eV.

| | 20eV | | | | 70eV | | | |
|---|---|---|---|---|---|---|---|---|
| | $\chi_{tot}$ = 2.22e-15 m³s⁻¹ | | $\chi^{ex}_{tot}$ = 4.94e-16 m³s⁻¹ | | $\chi_{tot}$ = 1.48e-14 m³s⁻¹ | | $\chi^{ex}_{tot}$ = 4.30e-15 m³s⁻¹ | |
| C III Level | $\chi_{III}$ (m³s⁻¹) | $\chi_{III}/\chi_{tot}$ | $\chi^{ex}_{III}$ (m³s⁻¹) | $\chi^{ex}_{III}/\chi^{ex}_{tot}$ | $\chi_{III}$ (m³s⁻¹) | $\chi_{III}/\chi_{tot}$ | $\chi^{ex}_{III}$ (m³s⁻¹) | $\chi^{ex}_{III}/\chi^{ex}_{tot}$ |
| $2s^2\ ^1S_0$ | 5.69e-16 | 2.6e-1 | 3.56e-17 | 7.2e-2 | 3.94e-15 | 2.7e-1 | 3.43e-16 | 8.0e-2 |
| $2s2p\ ^3P_0$ | 1.85e-16 | 8.3e-2 | 5.08e-17 | 1.0e-1 | 1.22e-15 | 8.2e-2 | 4.40e-16 | 1.0e-1 |
| $2s2p\ ^3P_1$ | 5.50e-16 | 2.5e-1 | 1.52e-16 | 3.1e-1 | 3.63e-15 | 2.5e-1 | 1.32e-15 | 3.1e-1 |
| $2s2p\ ^3P_2$ | 9.06e-16 | 4.1e-1 | 2.53e-16 | 5.1e-1 | 6.00e-15 | 4.1e-1 | 2.19e-15 | 5.1e-1 |
| $2s2p\ ^1P_1$ | 4.68e-19 | 2.1e-4 | 1.41e-19 | 2.9e-4 | 2.40e-18 | 1.6e-4 | 9.51e-19 | 2.2e-4 |
| $2p^2\ ^3P_0$ | 6.95e-20 | 3.1e-5 | 6.95e-20 | 1.4e-4 | 4.11e-19 | 2.8e-5 | 4.11e-19 | 9.6e-5 |
| $2p^2\ ^3P_1$ | 2.08e-19 | 9.4e-5 | 2.08e-19 | 4.2e-4 | 1.23e-18 | 8.3e-5 | 1.23e-18 | 2.9e-4 |
| $2p^2\ ^3P_2$ | 3.38e-19 | 1.5e-4 | 3.38e-19 | 6.8e-4 | 2.01e-18 | 1.4e-4 | 2.01e-18 | 4.7e-4 |
| $2p^2\ ^1D_2$ | 5.52e-19 | 2.5e-4 | 4.82e-19 | 9.8e-4 | 1.84e-18 | 1.2e-4 | 1.65e-18 | 3.8e-4 |
| $2p^2\ ^1S_0$ | 5.93e-21 | 2.7e-6 | 5.21e-21 | 1.1e-5 | 1.94e-20 | 1.3e-6 | 1.78e-20 | 4.1e-6 |
| $2s3s\ ^3S_1$ | 5.81e-22 | 2.6e-5 | 3.18e-21 | 6.4e-6 | 2.09e-19 | 1.4e-5 | 3.67e-20 | 8.5e-6 |
| $2s3s\ ^1S_0$ | 7.35e-20 | 3.3e-5 | 5.81e-21 | 1.2e-5 | 2.52e-19 | 1.7e-5 | 5.26e-20 | 1.2e-5 |
| $2s3p\ ^1P_1$ | 1.49e-20 | 6.7e-6 | 1.05e-21 | 2.1e-6 | 5.54e-20 | 3.7e-6 | 1.02e-20 | 2.4e-6 |
| $2s3p\ ^3P_0$ | 3.84e-19 | 1.7e-4 | 1.11e-20 | 2.2e-5 | 1.12e-18 | 7.6e-5 | 1.38e-19 | 3.2e-5 |
| $2s3p\ ^3P_1$ | 1.11e-18 | 5.0e-4 | 3.21e-20 | 6.5e-5 | 3.25e-18 | 2.2e-4 | 4.00e-19 | 9.3e-5 |
| $2s3p\ ^3P_2$ | 1.80e-18 | 8.1e-4 | 5.24e-20 | 1.1e-4 | 5.30e-18 | 3.6e-4 | 6.53e-19 | 1.5e-4 |
| $2s3d\ ^3D_1$ | 1.70e-20 | 7.6e-6 | 4.49e-22 | 9.1e-7 | 7.50e-20 | 5.1e-6 | 7.32e-21 | 1.7e-6 |
| $2s3d\ ^3D_2$ | 2.81e-20 | 1.3e-5 | 7.43e-22 | 1.5e-6 | 1.25e-19 | 8.4e-6 | 1.22e-20 | 2.8e-6 |
| $2s3d\ ^3D_3$ | 3.85e-20 | 1.7e-5 | 1.02e-21 | 2.1e-6 | 1.72e-19 | 1.2e-5 | 1.70e-20 | 4.0e-6 |
| $2s3d\ ^1D_2$ | 1.67e-20 | 7.5e-6 | 5.07e-22 | 1.0e-6 | 6.75e-20 | 4.6e-6 | 7.38e-21 | 1.7e-6 |

Figure 2 shows ionization PECs for a number of C IV transitions, while table 6 lists their values and those of the corresponding excitation PECs at temperatures of 20 and 70 eV, together with the ionization / excitation PEC ratio. The wavelengths in this table are taken from the NIST compilation (Ralchenko *et al.* 2011). As may be seen from equation (1), the ratio $n_{g-1}/n_g$ is still required to determine the contribution of ionization to the level populations. From section 4d, this ratio is typically ~1 for C III and C IV in plasmas that are in equilibrium, but can be significantly larger during impurity influxes. The recombination PECs have not been included in the present analysis, this being appropriate for an ionizing plasma. In fact, the inclusion of recombination only affects the $1s^2 3p\ ^3P$ levels, providing additional contributions to the excitation PECs of the 312.4 Å transitions of ~12%.

From table 6, it can be seen that even in steady-state conditions, when $n_{g-1}/n_g$ ~1, ionization is beginning to affect the C IV level populations, with the $\varepsilon^{ion}/\varepsilon^{exc}$ ratio for the 1548.2 Å transition (1-3) being greater than 10%. During impurity influxes, when $n_{g-1}/n_g$ >1 the effect of the ionization on the $1s^2 2p$ (2 and 3), $1s^2 3s$ (4) and $1s^2 3p$ (5 and 6) level populations, in particular, becomes more appreciable. Figure 3 illustrates the ionization / excitation PEC ratio for transitions from these levels for three values of $n_{g-1}/n_g$ as a function of $T_e$. In both figures 2 and 3, the PECs belonging to lines within the same multiplet have been combined, both for clarity and because in most cases lines within a multiplet are not resolved experimentally. The exception is for the 1548.2 and 1550.8 Å transitions, which often can be resolved and for which the $\varepsilon^{ion}/\varepsilon^{exc}$ ratios are substantially different. It perhaps should be emphasized that although the data that are presented are for UV spectral lines, the same analysis applies to visible transitions. The dash-double dot line in figure 3 applies both to the 312.42 and 312.45 Å VUV lines and to two of the most important C IV visible features at 5801.3 and 5812.0 Å, since the latter have the same upper levels.

Table 6. Ionization and excitation PECs and their ratios for a number of important C IV transitions at $T_e$ = 20 and 70 eV.

| Level indices | Wavelength (Å) | 20eV | | | 70eV | | |
|---|---|---|---|---|---|---|---|
| | | $\varepsilon_{ij}^{ion}$ (m$^3$s$^{-1}$) | $\varepsilon_{ij}^{exc}$ (m$^3$s$^{-1}$) | $\varepsilon_{ij}^{ion}/\varepsilon_{ij}^{exc}$ | $\varepsilon_{ij}^{ion}$ (m$^3$s$^{-1}$) | $\varepsilon_{ij}^{exc}$ (m$^3$s$^{-1}$) | $\varepsilon_{ij}^{ion}/\varepsilon_{ij}^{exc}$ |
| 1 - 11 | 244.903 | 2.39e-20 | 4.07e-17 | 5.86e-4 | 9.87e-19 | 1.45e-16 | 6.89e-3 |
| 1 - 10 | 244.911 | 1.20e-20 | 2.03e-17 | 5.91e-4 | 5.01e-19 | 7.23e-17 | 6.93e-3 |
| 2 - 12 | 289.141 | 3.65e-21 | 2.94e-17 | 1.24e-4 | 1.54e-19 | 1.08e-16 | 1.43e-3 |
| 3 - 13 | 289.228 | 6.63e-21 | 5.30e-17 | 1.25e-4 | 2.78e-19 | 1.94e-16 | 1.43e-3 |
| 3 - 12 | 289.231 | 7.30e-22 | 5.88e-18 | 1.24e-4 | 3.08e-20 | 2.16e-17 | 1.43e-3 |
| 2 - 9 | 296.856 | 8.94e-21 | 1.28e-17 | 6.98e-4 | 3.90e-19 | 4.11e-17 | 9.48e-3 |
| 3 - 9 | 296.951 | 1.79e-20 | 2.56e-17 | 6.98e-4 | 7.80e-19 | 8.20e-17 | 9.50e-3 |
| 1 - 6 | 312.420 | 4.99e-19 | 2.50e-16 | 2.00e-3 | 1.39e-17 | 7.29e-16 | 1.90e-2 |
| 1 - 5 | 312.451 | 2.51e-19 | 1.24e-16 | 2.02e-3 | 6.96e-18 | 3.63e-16 | 1.92e-2 |
| 2 - 7 | 384.031 | 6.80e-20 | 2.57e-16 | 2.64e-4 | 1.94e-18 | 7.41e-16 | 2.61e-3 |
| 3 - 8 | 384.174 | 1.23e-19 | 4.65e-16 | 2.65e-4 | 3.49e-18 | 1.34e-15 | 2.61e-3 |
| 3 - 7 | 384.190 | 1.36e-20 | 5.15e-17 | 2.64e-4 | 3.87e-19 | 1.49e-16 | 2.60e-3 |
| 2 - 4 | 419.525 | 3.07e-19 | 1.88e-16 | 1.63e-3 | 8.41e-18 | 4.07e-16 | 2.07e-2 |
| 3 - 4 | 419.714 | 6.14e-19 | 3.75e-16 | 1.64e-3 | 1.68e-17 | 8.11e-16 | 2.08e-2 |
| 7 - 14 | 1168.847 | 3.48e-22 | 2.39e-17 | 1.46e-5 | 9.63e-21 | 6.63e-17 | 1.45e-4 |
| 8 - 15 | 1168.990 | 4.91e-22 | 3.43e-17 | 1.43e-5 | 1.36e-20 | 9.49e-17 | 1.43e-4 |
| 8 - 14 | 1168.990 | 2.48e-23 | 1.71e-18 | 1.45e-5 | 6.88e-22 | 4.74e-18 | 1.45e-4 |
| 1 - 3 | 1548.187 | 3.27e-16 | 2.20e-14 | 1.47e-2 | 2.83e-15 | 2.09e-14 | 1.35e-1 |
| 1 - 2 | 1550.772 | 1.64e-16 | 4.43e-14 | 3.71e-3 | 1.42e-15 | 4.17e-14 | 3.40e-2 |

From table 5, it appears that the only important initial C III levels for the ionization are the ground and metastable ones and that ionization from the higher C III levels is orders of magnitude lower. However, when looking at the detailed ionization channels to particular C IV levels, it is found that this is a simplification. To illustrate this point the transitions with the highest ionization/excitation PEC ratios have been chosen and the fractional contributions to their upper levels, $\chi_{ki}/\chi_{IV}$, where

$$\chi_{ki} = \frac{n_k}{n_{g-1}} s_{ki},$$

are listed in table 7 at an electron temperature of 20 eV. The fractional contributions do not change by more than a factor of ~2.5 over the temperature range of interest and hence a single temperature illustrates the relative importance of the different ionizing channels. Only those C III levels which give rise to a fractional contribution greater than ~5×10$^{-3}$ for at least one C IV level are included in table 7. For all of these it can be seen that two or more of the C III metastable levels dominate the ionization. In four of the cases ionization from these levels is much more important than from the ground state. Furthermore, ionization from higher C III levels, although generally small is not necessarily negligible, in some cases exceeding ionization from the ground state. More accurate calculations of ionization rate coefficients are desirable to confirm these conclusions and to take account of possible resonances which may have a significant effect on the populating channels.

Table 7. Fractional ionization contributions, $\chi_{ki}/\chi_{IV}$, from the most important C III populating levels to the C IV levels most affected by ionization at $T_e = 20$ eV.

|  | C IV level | | | | | |
|---|---|---|---|---|---|---|
|  | $1s^22p\ ^2P_{1/2}$ | $1s^22p\ ^2P_{3/2}$ | $1s^23s\ ^2S_{1/2}$ | $1s^23p\ ^2P_{1/2}$ | $1s^23p\ ^2P_{3/2}$ | $1s^24s\ ^2S_{1/2}$ |
| $\chi_{IV}$ (m$^3$s$^{-1}$) | 1.64e-16 | 3.28e-16 | 9.22e-19 | 2.53e-19 | 5.03e-19 | 4.53e-20 |
| C III level | $\chi_{ki}/\chi_{IV}$ | $\chi_{ki}/\chi_{IV}$ | $\chi_{ki}/\chi_{IV}$ | $\chi_{ki}/\chi_{IV}$ | $\chi_{ki}/\chi_{IV}$ | $\chi_{ki}/\chi_{IV}$ |
| $1s^22s^2\ ^1S_0$ | 7.3e-2 | 7.1e-2 | 3.3e-1 | 1.6e-2 | 1.6e-2 | 3.2e-1 |
| $1s^22s2p\ ^3P_0$ | 3.1e-1 | 1.4e-3 | 7.5e-2 | 2.9e-1 | 7.5e-5 | 7.5e-2 |
| $1s^22s2p\ ^3P_1$ | 6.1e-1 | 1.6e-1 | 2.2e-1 | 5.7e-1 | 1.4e-1 | 2.2e-1 |
| $1s^22s2p\ ^3P_2$ | 6.8e-3 | 7.7e-1 | 3.6e-1 | 3.6e-4 | 7.2e-1 | 3.7e-1 |
| $1s^22s3p\ ^3P_0$ | 4.4e-6 | 7.6e-6 | 1.9e-5 | 2.9e-2 | 1.0e-6 | 9.8e-5 |
| $1s^22s3p\ ^3P_1$ | 1.6e-5 | 2.0e-5 | 5.6e-5 | 5.6e-2 | 1.4e-2 | 2.8e-4 |
| $1s^22s3p\ ^3P_2$ | 3.5e-5 | 2.8e-5 | 9.0e-5 | 4.8e-6 | 6.9e-2 | 4.5e-4 |
| $1s^22s4s\ ^1S_0$ | 1.6e-6 | 1.6e-6 | 2.7e-5 | 5.0e-5 | 4.9e-5 | 4.8e-3 |
| $1s^22p3p\ ^3D_1$ | 4.7e-4 | 5.1e-5 | 8.2e-7 | 1.7e-2 | 1.8e-3 | 1.0e-6 |
| $1s^22p3p\ ^3D_2$ | 4.8e-4 | 2.4e-4 | 1.4e-6 | 1.7e-2 | 8.9e-3 | 1.6e-6 |
| $1s^22p3p\ ^3D_3$ | 5.5e-6 | 6.6e-4 | 1.8e-6 | 4.2e-7 | 2.4e-2 | 2.2e-6 |

**b. C V**

A similar analysis has been carried out for the ionization of C IV to investigate its effect on the C V excited level populations. As before, FAC calculations were undertaken, all configurations up to and including those in the $n = 5$ shell being included for the initial C IV stage, with the $n = 1$ shell being taken to be closed. For C V the ground configuration, $1s^2$, and excited configurations $mlpl'$, where $m = 1, 2$ and $p = 2, 3, 4, 5$, were included. Again calculations with all three FAC ionization rate options were performed. However, in the following tables the results are those derived using the distorted-wave option. All calculations have been performed for an electron density of $10^{19}$ m$^{-3}$.

In table 8 the ionization contributions from all C IV levels, $k$,

$$\chi_V = \sum_k \frac{n_k}{n_{g-1}} s_{ki},$$

to the lowest 31 individual C V levels, $i$, and their fractions of the total C IV to C V ionization contributions are listed at electron temperatures of 70 and 130 eV. The former is the highest temperature at which C IV is expected under conditions in which equilibrium is maintained, while 130 eV is the largest value of $T_e$ for which the population models could be calculated without extrapolation of the atomic data. This was determined by the range of temperatures for which the C IV R-matrix electron collisional excitation data are available. The latter temperature is, in any case, only expected to be reached in exceptional circumstances, such as during a C influx when the C ions flow into higher temperature plasma regions than would be encountered under equilibrium conditions.

Table 8. Ionization contributions to the individual C V levels, $\chi_V$, and their fractions of the total ionization contributions, $\chi_{tot}$, together with populations of the C V levels as a fraction of the ground level population, $n_g$, at $T_e = 70$ and $130$ eV.

| C V Level index | C V Level | 70eV $\chi_{tot} = 8.55\text{e-}16 \text{ m}^3\text{s}^{-1}$ | | | 130eV $\chi_{tot} = 1.45\text{e-}15 \text{ m}^3\text{s}^{-1}$ | | |
|---|---|---|---|---|---|---|---|
| | | $\chi_V$ (m$^3$s$^{-1}$) | $\chi_V / \chi_{tot}$ | $n_i / n_g$ | $\chi_V$ (m$^3$s$^{-1}$) | $\chi_V / \chi_{tot}$ | $n_i / n_g$ |
| 1 | 1s$^2$ $^1$S$_0$ | 8.54e-16 | 9.98e-1 | 1.00+0 | 1.42e-15 | 9.85e-1 | 1.00e+0 |
| 2 | 1s2s $^3$S$_1$ | 1.15e-18 | 1.3e-3 | 1.02e-3 | 1.60e-17 | 1.1e-2 | 4.48e-3 |
| 3 | 1s2s $^1$S$_0$ | 3.57e-19 | 4.2e-4 | 3.54e-5 | 5.13e-18 | 3.5e-3 | 1.84e-4 |
| 4 | 1s2p $^3$P$_1$ | 8.67e-22 | 1.0e-6 | 2.23e-6 | 1.00e-20 | 6.9e-6 | 8.02e-6 |
| 5 | 1s2p $^3$P$_0$ | 2.90e-22 | 3.4e-7 | 1.09e-6 | 3.35e-21 | 2.3e-6 | 3.90e-6 |
| 6 | 1s2p $^3$P$_2$ | 1.44e-21 | 1.7e-6 | 5.37e-6 | 1.67e-20 | 1.2e-5 | 1.93e-5 |
| 7 | 1s2p $^1$P$_1$ | 8.28e-22 | 9.7e-7 | 9.21e-11 | 9.83e-21 | 6.8e-6 | 4.26e-10 |
| 8 | 1s3s $^3$S$_1$ | 1.34e-20 | 1.6e-5 | 1.39e-9 | 2.63e-19 | 1.8e-4 | 6.08e-9 |
| 9 | 1s3s $^1$S$_0$ | 1.98e-21 | 2.3e-6 | 3.70e-10 | 3.95e-20 | 2.7e-5 | 1.94e-9 |
| 10 | 1s3p $^3$P$_1$ | 2.14e-23 | 2.5e-8 | 3.57e-10 | 3.41e-22 | 2.4e-7 | 1.76e-9 |
| 11 | 1s3p $^3$P$_0$ | 7.13e-24 | 8.3e-9 | 1.19e-10 | 1.14e-22 | 7.9e-8 | 5.87e-10 |
| 12 | 1s3p $^3$P$_2$ | 3.54e-23 | 4.1e-8 | 5.95e-10 | 5.65e-22 | 3.9e-7 | 2.93e-9 |
| 13 | 1s3d $^3$D$_1$ | 2.27e-25 | 2.7e-10 | 6.96e-11 | 3.51e-24 | 2.4e-9 | 3.28e-10 |
| 14 | 1s3d $^3$D$_2$ | 3.77e-25 | 4.4e-10 | 1.14e-10 | 5.84e-24 | 4.0e-9 | 5.35e-10 |
| 15 | 1s3d $^3$D$_3$ | 5.25e-25 | 6.1e-10 | 1.62e-10 | 8.14e-24 | 5.6e-9 | 7.63e-10 |
| 16 | 1s3d $^1$D$_2$ | 3.76e-25 | 4.4e-10 | 4.99e-11 | 5.83e-24 | 4.0e-9 | 2.05e-10 |
| 17 | 1s3p $^1$P$_1$ | 1.38e-23 | 1.6e-8 | 2.17e-11 | 2.22e-22 | 1.5e-7 | 1.28e-10 |
| 18 | 1s4s $^3$S$_1$ | 9.20e-22 | 1.1e-6 | 4.04e-10 | 2.04e-20 | 1.4e-5 | 1.90e-9 |
| 19 | 1s4s $^1$S$_0$ | 1.47e-22 | 1.7e-7 | 1.53e-10 | 3.28e-21 | 2.3e-6 | 8.58e-10 |
| 20 | 1s4p $^3$P$_1$ | 1.50e-24 | 1.8e-9 | 1.55e-10 | 2.66e-23 | 1.8e-8 | 7.85e-10 |
| 21 | 1s4p $^3$P$_0$ | 4.96e-25 | 5.8e-10 | 5.16e-11 | 8.81e-24 | 6.1e-9 | 2.62e-10 |
| 22 | 1s4p $^3$P$_2$ | 2.53e-24 | 3.0e-9 | 2.58e-10 | 4.48e-23 | 3.1e-8 | 1.31e-9 |
| 23 | 1s4d $^3$D$_1$ | 1.01e-25 | 1.2e-10 | 2.78e-11 | 1.67e-24 | 1.2e-9 | 1.34e-10 |
| 24 | 1s4d $^3$D$_2$ | 1.65e-25 | 1.9e-10 | 4.58e-11 | 2.72e-24 | 1.9e-9 | 2.22e-10 |
| 25 | 1s4d $^3$D$_3$ | 2.20e-25 | 2.6e-10 | 6.48e-11 | 3.64e-24 | 2.5e-9 | 3.14e-10 |
| 26 | 1s4f $^1$F$_3$ | 1.33e-25 | 1.6e-10 | 2.73e-11 | 1.96e-24 | 1.4e-9 | 1.07e-10 |
| 27 | 1s4f $^3$F$_3$ | 1.33e-25 | 1.6e-10 | 3.52e-11 | 1.96e-24 | 1.4e-9 | 1.51e-10 |
| 28 | 1s4f $^3$F$_4$ | 1.71e-25 | 2.0e-10 | 5.26e-11 | 2.52e-24 | 1.7e-9 | 2.35e-10 |
| 29 | 1s4f $^3$F$_2$ | 9.48e-26 | 1.1e-10 | 2.92e-11 | 1.40e-24 | 9.7e-10 | 1.30e-10 |
| 30 | 1s4d $^1$D$_2$ | 1.61e-25 | 1.9e-10 | 3.66e-11 | 2.66e-24 | 1.8e-9 | 1.66e-10 |
| 31 | 1s4p $^1$P$_1$ | 1.12e-24 | 1.3e-9 | 1.28e-11 | 1.98e-23 | 1.4e-8 | 8.06e-11 |

It can be seen from table 8 that the distribution of contributions is much closer to that expected, with 99.7% and 98% of ionization terminating directly in the ground state at temperatures of 70 and 130 eV, respectively. This is not surprising given the large energy gap between the ground and first excited levels. In table 9, the total ionization contributions from different C IV levels,

$$\chi_{IV} = \sum_i \frac{n_k}{n_{g-1}} s_{ki},$$

are listed, together with their fraction of the total ionization contribution. The corresponding figures are also given for the ionization contributions to the C V excited levels alone,

$$\chi^{ex}_{IV} = \sum_{i>1} \frac{n_k}{n_{g-1}} s_{ki}.$$

In both cases the ionization is dominated by the contribution (99.8%) from the $1s^22s\ ^2S_{1/2}$ ground state. This is very much the expected 'ground state to ground state' scenario.

Table 9. Ionization contributions from the individual C IV levels to all C V levels, $\chi_{IV}$, and to the excited C V levels, $\chi^{ex}_{IV}$, and their fractions of the total ionization contributions at $T_e = 70$ and 130 eV.

| | 70eV | | | | 130eV | | | |
|---|---|---|---|---|---|---|---|---|
| | $\chi_{tot}$ = 8.55e-16 m$^3$s$^{-1}$ | | $\chi^{ex}_{tot}$ = 1.53e-18 m$^3$s$^{-1}$ | | $\chi_{tot}$ = 1.45e-15 m$^3$s$^{-1}$ | | $\chi^{ex}_{tot}$ = 2.14e-17 m$^3$s$^{-1}$ | |
| C IV Level | $\chi_{IV}$ (m$^3$s$^{-1}$) | $\chi_{IV}/\chi_{tot}$ | $\chi^{ex}_{IV}$ (m$^3$s$^{-1}$) | $\chi^{ex}_{IV}/\chi^{ex}_{tot}$ | $\chi_{IV}$ (m$^3$s$^{-1}$) | $\chi_{IV}/\chi_{tot}$ | $\chi^{ex}_{IV}$ (m$^3$s$^{-1}$) | $\chi^{ex}_{IV}/\chi^{ex}_{tot}$ |
| 2s $^2S_{1/2}$ | 8.52e-16 | 9.96e-1 | 1.53e-18 | 9.98e-1 | 1.44e-15 | 9.97e-1 | 2.14e-17 | 9.98e-1 |
| 2p $^2P_{1/2}$ | 1.03e-18 | 1.2e-3 | 1.17e-21 | 7.7e-4 | 1.34e-18 | 9.3e-4 | 1.38e-20 | 6.0e-4 |
| 2p $^2P_{3/2}$ | 2.01e-18 | 2.3e-3 | 2.33e-21 | 1.5e-3 | 2.64e-18 | 1.8e-3 | 2.74e-20 | 1.2e-3 |
| 3s $^2S_{1/2}$ | 1.12e-20 | 1.3e-5 | 3.56e-24 | 2.3e-6 | 1.27e-20 | 8.8e-6 | 5.09e-23 | 2.3e-6 |
| 3p $^2P_{1/2}$ | 4.09e-21 | 4.8e-6 | 9.90e-25 | 6.5e-7 | 5.13e-21 | 3.6e-6 | 1.64e-23 | 7.6e-7 |
| 3p $^2P_{3/2}$ | 8.19e-21 | 9.6e-6 | 1.98e-24 | 1.3e-6 | 1.03e-20 | 7.1e-6 | 3.29e-23 | 1.5e-6 |
| 3d $^2D_{3/2}$ | 3.83e-21 | 4.5e-6 | 6.13e-25 | 4.0e-7 | 4.51e-21 | 3.1e-6 | 9.59e-24 | 4.4e-7 |
| 3d $^2D_{5/2}$ | 5.67e-21 | 6.6e-6 | 9.21e-25 | 6.0e-7 | 6.70e-21 | 4.6e-6 | 1.44e-23 | 6.6e-7 |
| 4s $^2S_{1/2}$ | 7.09e-21 | 8.3e-6 | 8.16e-25 | 5.3e-7 | 6.69e-21 | 4.6e-6 | 1.27e-23 | 5.8e-7 |
| 4p $^2P_{1/2}$ | 9.64e-22 | 1.1e-6 | 1.16e-25 | 7.6e-8 | 1.10e-21 | 7.6e-7 | 1.95e-24 | 9.1e-8 |
| 4p $^2P_{3/2}$ | 5.17e-21 | 6.0e-6 | 6.20e-25 | 4.1e-7 | 5.87e-21 | 4.1e-6 | 1.04e-23 | 4.8e-7 |
| 4d $^2D_{3/2}$ | 3.06e-21 | 3.6e-6 | 2.20e-25 | 1.4e-7 | 3.09e-21 | 2.1e-6 | 3.59e-24 | 1.6e-7 |
| 4d $^2D_{5/2}$ | 5.11e-21 | 6.0e-6 | 3.30e-25 | 2.2e-7 | 4.54e-21 | 3.1e-6 | 5.39e-24 | 2.5e-7 |
| 4f $^2F_{5/2}$ | 3.78e-21 | 4.4e-6 | 2.12e-25 | 1.4e-7 | 3.68e-21 | 2.5e-6 | 3.13e-24 | 1.4e-7 |
| 4f $^2F_{7/2}$ | 5.05e-21 | 5.9e-6 | 2.83e-25 | 1.9e-7 | 4.92e-21 | 3.4e-6 | 4.18e-24 | 1.9e-7 |

Despite the ionization contributions to the excited C V levels being very small, from table 8 it can be seen that many of the C V populations are also small. For example, that for 1s2p $^1P_1$ in the second excited configuration is $(1 - 4) \times 10^{-10} n_g$ at the temperatures listed, due to its fast decay to the ground state. Hence, it is thought worthwhile to check the ionization PECs and their ratio to the excitation PECs. The C IV population calculation used to generate the ionization PECs excludes charge exchange recombination, although we note that it makes little difference to the PECs (at most ~1%). Table 10 lists both ionization and excitation PECs for a number of transitions, while figure 4 shows the ionization PECs as a function of $T_e$ for the transitions with the largest ionization / excitation PEC ratios. Examples of transitions from all apart from level 3 (1s2s $^1S_0$) of the lowest 19 C V levels are represented in the table, as there is no observed radiative transition from level 3. A number of the ionization / excitation PEC ratios are unexpectedly large, the highest (~7%) involving transitions from the 1s3s $^3S_1$ level. Figure 5 illustrates the ionization / excitation PEC ratio for the transitions with the largest ratios for three values of $n_{g-1}/n_g$ as a function of $T_e$. As discussed in section 4d, C V ions have a wide spatial extent, resulting in a low equilibrium value of $n_{g-1}/n_g$ (typically ~0.2-0.5) when using line integrated measurements. This offsets, to some extent, the high $\varepsilon^{ion}_{ij}/\varepsilon^{exc}_{ij}$ ratios. However, if spatially resolved measurements are possible, then $n_{g-1}/n_g$ will be ~1 in the lower temperature, outer regions of the C V emission shell. Unlike the C III ionization, which even in steady-state is beginning to have a significant effect on the C IV excited level populations, it can be seen that, under these conditions, the effect of the C IV ionization on the C V excited level populations is marginal. However, during impurity influxes, when $n_{g-1} / n_g > 1$, these results show that the populations of excited states will be affected by ionization and more accurate calculations of ionization rate coefficients are therefore important.

Table 10. Ionization and excitation PECs and their ratios for the most important C V transitions at $T_e$ = 70 and 130 eV.

| Transition indices | Wavelength (Å) | 70eV | | | 130eV | | |
|---|---|---|---|---|---|---|---|
| | | $\varepsilon_{ij}^{ion}$ (m$^3$s$^{-1}$) | $\varepsilon_{ij}^{exc}$ (m$^3$s$^{-1}$) | $\varepsilon_{ij}^{ion} / \varepsilon_{ij}^{exc}$ | $\varepsilon_{ij}^{ion}$ (m$^3$s$^{-1}$) | $\varepsilon_{ij}^{exc}$ (m$^3$s$^{-1}$) | $\varepsilon_{ij}^{ion} / \varepsilon_{ij}^{exc}$ |
| 1 - 17 | 34.973 | 1.33e-23 | 6.93e-19 | 1.92e-5 | 2.13e-22 | 4.08e-18 | 5.22e-5 |
| 1 - 10 | 35.070 | 1.46e-26 | 3.20e-22 | 4.55e-5 | 2.33e-25 | 1.58e-21 | 1.47e-4 |
| 1 - 7 | 40.268 | 8.28e-22 | 8.79e-18 | 9.41e-5 | 9.83e-21 | 4.05e-17 | 2.43e-4 |
| 1 - 4 | 40.731 | 2.74e-22 | 6.20e-18 | 4.42e-5 | 3.17e-21 | 2.23e-17 | 1.42e-4 |
| 1 - 2 | 41.472 | 1.15e-18 | 4.27e-21 | metastable | 1.60e-17 | 1.88e-20 | metastable |
| 4 - 18 | 189.255 | 1.69e-22 | 2.90e-20 | 5.83e-3 | 3.75e-21 | 1.36e-19 | 2.75e-2 |
| 5 - 18 | 189.260 | 5.67e-23 | 9.72e-21 | 5.83e-3 | 1.25e-21 | 4.56e-20 | 2.75e-2 |
| 6 - 18 | 189.304 | 2.83e-22 | 4.85e-20 | 5.84e-3 | 6.26e-21 | 2.27e-19 | 2.75e-2 |
| 7 - 19 | 198.069 | 9.94e-23 | 5.37e-20 | 1.85e-3 | 2.21e-21 | 3.02e-19 | 7.34e-3 |
| 2 - 12 | 227.182 | 3.54e-23 | 7.81e-19 | 4.54e-5 | 5.65e-22 | 3.85e-18 | 1.47e-4 |
| 2 - 11 | 227.202 | 7.13e-24 | 1.57e-19 | 4.53e-5 | 1.14e-22 | 7.71e-19 | 1.47e-4 |
| 2 - 10 | 227.203 | 2.13e-23 | 4.69e-19 | 4.55e-5 | 3.40e-22 | 2.31e-18 | 1.47e-4 |
| 3 - 17 | 247.315 | 5.67e-25 | 2.96e-20 | 1.92e-5 | 9.10e-24 | 1.74e-19 | 5.22e-5 |
| 4 - 14 | 248.664 | 2.82e-25 | 3.62e-19 | 7.81e-7 | 4.37e-24 | 1.69e-18 | 2.58e-6 |
| 4 - 13 | 248.664 | 9.45e-26 | 1.24e-19 | 7.61e-7 | 1.46e-24 | 5.84e-19 | 2.51e-6 |
| 5 - 13 | 248.672 | 1.26e-25 | 1.66e-19 | 7.60e-7 | 1.95e-24 | 7.79e-19 | 2.51e-6 |
| 6 - 15 | 248.741 | 5.26e-25 | 6.87e-19 | 7.66e-7 | 8.15e-24 | 3.23e-18 | 2.52e-6 |
| 6 - 14 | 248.748 | 9.40e-26 | 1.20e-19 | 7.81e-7 | 1.46e-24 | 5.64e-19 | 2.58e-6 |
| 6 - 13 | 248.748 | 6.30e-27 | 8.28e-21 | 7.61e-7 | 9.76e-26 | 3.89e-20 | 2.51e-6 |
| 4 - 8 | 260.135 | 4.45e-21 | 2.93e-19 | 1.52e-2 | 8.77e-20 | 1.28e-18 | 6.84e-2 |
| 5 - 8 | 260.143 | 1.49e-21 | 9.78e-20 | 1.52e-2 | 2.93e-20 | 4.28e-19 | 6.84e-2 |
| 6 - 8 | 260.227 | 7.43e-21 | 4.89e-19 | 1.52e-2 | 1.46e-19 | 2.14e-18 | 6.85e-2 |
| 7 - 16 | 267.267 | 3.75e-25 | 1.97e-19 | 1.90e-6 | 5.81e-24 | 8.12e-19 | 7.16e-6 |
| 7 - 14 | 267.427 | 9.35e-28 | 1.20e-21 | 7.81e-7 | 1.45e-26 | 5.62e-21 | 2.58e-6 |
| 7 - 9 | 271.882 | 1.98e-21 | 2.35e-19 | 8.44e-3 | 3.95e-20 | 1.23e-18 | 3.21e-2 |
| 10 - 18 | 756.779 | 1.37e-22 | 2.35e-20 | 5.84e-3 | 3.03e-21 | 1.10e-19 | 2.75e-2 |
| 11 - 18 | 756.788 | 4.57e-23 | 7.83e-21 | 5.83e-3 | 1.01e-21 | 3.68e-20 | 2.75e-2 |
| 12 - 18 | 757.009 | 2.29e-22 | 3.92e-20 | 5.84e-3 | 5.06e-21 | 1.84e-19 | 2.75e-2 |
| 17 - 19 | 775.930 | 4.79e-23 | 2.59e-20 | 1.85e-3 | 1.07e-21 | 1.45e-19 | 7.34e-3 |
| 2 - 6 | 2270.89 | 1.44e-21 | 3.26e-17 | 4.41e-5 | 1.67e-20 | 1.17e-16 | 1.42e-4 |
| 2 - 5 | 2277.27 | 2.90e-22 | 6.54e-18 | 4.43e-5 | 3.35e-21 | 2.35e-17 | 1.43e-4 |
| 2 - 4 | 2277.92 | 5.94e-22 | 1.34e-17 | 4.42e-5 | 6.87e-21 | 4.83e-17 | 1.42e-4 |
| 3 - 7 | 3526.67 | 1.13e-26 | 1.20e-22 | 9.41e-5 | 1.35e-25 | 5.55e-22 | 2.43e-4 |

Finally, table 11 lists the contributions from the different C IV levels to the most affected C V levels at a temperature of 70 eV. Here the situation tends to be simpler than for the C III to C IV ionization, with the C V levels from which transitions with the highest $\varepsilon_{ij}^{ion} / \varepsilon_{ij}^{exc}$ ratio originate being populated almost entirely from the C IV ground state. In some other cases, for example, the 1s2p $^1P_1$ level illustrated in the table, there is no contribution from the ground level and ionization from, in this case, the 2p $^2P$ levels are the dominant channels. The 3d $^2D$ and 4f $^2F$ levels play similarly dominant roles. Nevertheless, when this happens the resulting populations and, hence, the line intensities do not appear to be significantly affected by ionization.

Table 11. Fractional ionization contributions, $\chi_{ki} / \chi_V$, from the most important C IV populating levels to the C V levels most affected by ionization at $T_e = 70$ eV.

|  | C V level | | | | |
|---|---|---|---|---|---|
|  | 1s2p $^1P_1$ | 1s3s $^3S_1$ | 1s3s $^1S_0$ | 1s4s $^3S_1$ | 1s4s $^1S_0$ |
| $\chi_V$ (m$^3$s$^{-1}$) | 8.28e-22 | 1.34e-20 | 1.98e-21 | 9.20e-22 | 1.47e-22 |
| C IV level | $\chi_{ki} / \chi_V$ | $\chi_{ki} / \chi_V$ | $\chi_{ki} / \chi_V$ | $\chi_{ki} / \chi_V$ | $\chi_{ki} / \chi_V$ |
| 1s$^2$2s $^2S_{1/2}$ | 0.0 | 9.998e-1 | 9.996e-1 | 9.992e-1 | 9.98e-1 |
| 1s$^2$2p $^2P_{1/2}$ | 3.3e-1 | 0.0 | 3.9e-7 | 0.0 | 1.0e-6 |
| 1s$^2$2p $^2P_{3/2}$ | 6.7e-1 | 0.0 | 7.4e-7 | 0.0 | 1.9e-6 |
| 1s$^2$3s $^2S_{1/2}$ | 0.0 | 1.8e-4 | 4.1e-4 | 1.7e-4 | 2.4e-4 |
| 1s$^2$3p $^2P_{3/2}$ | 2.8e-5 | 0.0 | 1.2e-10 | 0.0 | 8.4e-11 |
| 1s$^2$4s $^2S_{1/2}$ | 0.0 | 3.6e-6 | 5.8e-6 | 6.0e-4 | 1.3e-3 |

## 6. Discussion

The results presented in sections 3 and 5 provide no evidence of ionization from C IV and C V excited levels significantly altering their populations, but show that the populations can be affected when the excited levels are the final state of the ionization process. Although in steady-state plasmas the latter effect is on the whole marginal, an exception being to the C IV 1s$^2$2p levels, during transient events such as impurity influxes this mechanism is expected to play a significant role in determining the C IV and C V excited level populations. It is instructive to compare the present results with a study by Loch *et al.* (2004). These authors compare two codes using collisional-radiative models, ADAS (Summers, 2004) and the Los Alamos National Laboratory (LANL) suite of codes (Abdallah *et al.* 1994) by applying them to the three ionization stages of Li. ADAS is most suited to low density laboratory and astrophysical plasmas and provides flexibility in the atomic data that can be employed; the LANL codes generate their own distorted-wave atomic datasets. This comparison allows an assessment of not only the codes, but also the atomic data used, highlighting the importance of the different mechanisms involved in the collisional-radiative models. In particular, the models are applied to calculating the ionization balance and total radiated power loss. In contrast, the present study concentrates on the detailed calculation of the population of the so-called 'spectroscopic levels', i.e. those within the $n \leq 5$ shells, which are of most importance for diagnostic measurements. Since Loch *et al.* include high-lying energy levels in their analysis, for which $n >> 5$, they come to a different conclusion regarding ionization from excited levels. They find that such ionization can affect the excited state populations. This is because with increasing $n$ the radiative transition probabilities tend to decrease, while there is a significant increase in the ionization rate coefficients until ionization becomes a significant depopulating mechanism. These high-lying levels have very small populations and, although they can provide a channel that influences the ionization balance, individual lines from these levels cannot be observed or resolved with the usual spectrometers employed on laboratory devices. They are therefore not considered in the present study.

Loch *et al.* (2004) also make the assumption that ionization to and recombination from excited levels can be neglected at the low densities being studied ($n_e \leq 10^{20}$ m$^{-3}$). They justify this approximation by noting the good agreement obtained for the ionization balance between the ADAS and LANL codes; the former does not include these channels, while the latter does. Although for their purpose of deriving the ionization balance, this approximation is usually valid, it cannot be generalized to include the detailed population calculations for the spectroscopic levels. For example, our table 8 shows the very small contribution of ionization to the non-metastable excited states ($\leq 2\times 10^{-4}$ of the total) and, consequently, this has no effect on the ionization balance. In this case, the ionization balance is not a sensitive test of the approximation in general. An exception where ionization to non-metastable excited levels could affect even the ionization balance is C III to C IV, where table 4 shows the much higher contributions, in particular, to the $1s^2 2p$ $^2P_{1/2,3/2}$ levels of C IV (~10% and 20% of the total, respectively). If the ionization balance is calculated on the basis of the ionization channels ending only in the C IV ground state, then the rate will be underestimated. Again this emphasizes that the approximation cannot be generalized and would suggest that even when calculating the ionization balance, it needs to be assessed on a case-by-case basis.

Another important consideration is the accuracy of the atomic data used in the present study. FAC contains three different ionization calculations and significant differences are found among them, resulting in larger uncertainties than those due to the population models. Differences between the ionization calculations are typically up to a factor of 3, but in some cases can exceed this, with the distorted-wave rate coefficients tending to be higher than the CB and BED results. For the present purpose, variations in the ionization PECs are of more importance and these tend to be smaller since they are biased towards particular dominant contributions. Generally, the variation in the C IV PECs obtained using the different ionization calculations are a factor of ~2, whereas that for the C V PECs is smaller ( ~20%).

It is common in plasma modelling to use distorted-wave ionization calculations as in the present work. For example, the LANL codes rely entirely on these data. Nevertheless there is concern that, although the distorted-wave calculations are often satisfactory for ionization from the ground state, they may be less acceptable for ionization either beginning or terminating in excited states. The present FAC distorted-wave results for ionization from the C III ground state ($1s^2 2s^2$ $^1S_0$) are in good agreement ($\leq 6\%$) with the R-matrix with pseudostates calculation of Fogle *et al.* (2008), the latter yielding the best agreement with experiment of the various calculations reviewed. However, Griffin *et al.* (2005), one of a few papers to discuss ionization calculations from excited levels, find increasing discrepancies with increasing *n*. These authors compare perturbative distorted-wave calculations with benchmark non-perturbative R-matrix with pseudostates results for excited states in H-like ions. For example, distorted-wave calculations for ionization from the $n = 4$ levels of Li III lead to cross sections up to 50% larger than the R-matrix data. On the other hand, Pindzola *et al.* (2011), who also compare R-matrix with pseudostates cross sections with those from distorted-wave calculations, find much better agreement for ionization from the $1s^2 5s$ configuration of the moderately charged higher Z C IV ion. Although this configuration is not used in the present analysis, the FAC distorted-wave cross section can also be compared with their results. The agreement is somewhat poorer, the FAC cross sections being generally ~35 - 50% smaller, except near threshold where larger differences are found. Such differences would lead to the FAC ionization rate coefficient being up to a factor of 2 lower. Discrepancies of this magnitude do not negate the conclusions of the present study. The emphasis of the above and similar publications (e.g. Lee *et al.* 2010 who consider B I to III) is to obtain a satisfactory *n* scaling so that ionization from very high *n* levels, which affects the ionization balance, can be adequately described. However, there is also a clear need for the (computationally demanding) non-perturbative calculations, such as the R-matrix with pseudostates method for ionization *to* excited spectroscopic levels. This is not only to benchmark the direct ionization cross sections. Although the EA and REDA indirect ionization processes can be investigated with, for example, the isolated resonance distorted-wave approximation and close-coupling calculations, the R-matrix approach is required for a full treatment of all direct and indirect processes together with interference effects between the different channels.

## 7. Conclusions

A study has been made of the effect of direct electron collisional ionization on the C IV and C V excited energy level populations in laboratory plasmas for which $n_e < 10^{23}$ - $10^{24}$ m$^{-3}$. Ionization cross sections and rate coefficients from and to excited energy levels are scarce in the literature and so these data have been generated using the distorted-wave option of FAC (Gu 2003). For comparison, CB and BED calculations within FAC have also been undertaken. Ionization PECs have been derived and it is noted that the variations due to the different ionization calculations lead to much smaller changes in the ionization PECs than in the cross sections and rate coefficients, within a factor of 2 for C IV and ~20% for C V. The present analysis only considers the so-called 'spectroscopic levels' ($n \leq 5$), which are of most importance for diagnostic purposes in low density laboratory plasmas such as those found in tokamaks.

When assessing direct electron collisional ionization from C IV and C V excited levels, the calculated ionization rates are compared with radiative decay rates, this latter being the normal depopulating mechanism. It is found that for all excited levels ($n \leq 5$) except the two C V metastable ones, the ionization rates are negligible compared with the transition probabilities. In the case of the metastable levels, which have low radiative decay rates and where electron and heavy particle collisions are the dominant depopulating mechanisms, the ionization rates are still small, but much closer to the collisional excitation and de-excitation rates.

The importance of ionization to the C IV and C V excited levels is also investigated. In the case of C III ionization to excited C IV levels, the picture that emerges is different from the often assumed 'ground to ground state' scenario, with a significant proportion of the ionization (~10% and 20%) terminating in the first excited levels, $1s^2 2p$ $^2P_{1/2}$ and $^2P_{3/2}$. This is sufficiently large to affect ionization balance calculations. Although there is far less ionization to the more highly excited states, the population due to electron collisional excitation is also small, with the result that the direct collisional ionization has a much greater effect than might otherwise be expected.

To assess the effect, the analysis compares ionization and excitation PECs. The present calculations suggest that during steady-state operation, ionization is becoming sufficiently important to have a significant effect on the C IV $1s^2 2p$ $^2P_{3/2}$ level population. However, it is not able to explain the inconsistency in the modelling of the main chamber SOL impurities (Lawson *et al*. 2012), for which larger rate coefficients would be required. Nevertheless, ionization will play an even more important role in determining level populations during transient events, such as impurity influxes, when the populations of the ionization stages are distorted from their equilibrium values. Measurements of influxes suggest that the population of the initial ionization stage can increase by factors of up to ~5 relative to that of the final ionization stage and, in more extreme cases, increases of an order of magnitude are possible. That ionization is significant during impurity influxes might be expected from observations of the so-called 'ionization feature', which has an initial rapid decay and a later slower fall-off directly related to the impurity transport. Indeed, such transient events can be of particular use in impurity transport studies. The analysis is completed by indicating which C III levels dominate the ionization channels to the most affected C IV levels.

The ionization of C IV better fits the 'ground to ground state' scenario, with ~98% or more ionization terminating in the C V ground level. However, even in this case, ionization is more important than might be expected due to the very low populations in the C V excited states. The present calculations suggest that ionization will have only a marginal effect on the excited C V level populations during steady-state conditions, when equilibrium is maintained, but will be important during transient events, when the ionization stage populations increase, the initial ionization stage disproportionally so. Our analysis indicates that for those transitions most likely to be affected by ionization, the dominant ionization channel to their upper level originates in the C IV ground level ($1s^2 2s\ ^2S_{1/2}$), making this a priority for more accurate calculations. It follows that, just as charge exchange recombination needs to be taken into account in order to obtain accurate level populations in recombining plasmas, ionization must be included during ionizing events. Future work must include the calculation of ionization rate coefficients for Be, N and Ne, all important impurities in the JET plasmas, either intrinsic (Be) or gas-puffed (N and Ne) for radiative divertor studies.

A limitation of the present analysis is the use of distorted-wave calculations of the direct collisional ionization cross sections and ionization rate coefficients. There is clearly a need for more accurate non-pertubative calculations for the C III to C V ions, in part to benchmark the computationally fast FAC data, which would then allow a number of ionization stages and elements to be easily treated. R-matrix calculations would also allow the direct and indirect (EA, REDA and READI) processes together with interference between the different channels to be investigated fully. Expected differences in the results of the calculations should not alter the main conclusions of the present work, in particular that ionization does need to be considered as a populating mechanism of the excited so-called 'spectroscopic levels' ($n \leq 5$) during impurity influxes (in very accurate work even during steady-state). Indeed, the inclusion of the indirect ionization processes, which may preferentially populate excited levels would tend to increase further the importance of ionization to these levels.

## Acknowledgements

This work, supported by the European Communities under the contract of Association between EURATOM and CCFE, was carried out within the framework of the European Fusion Development Agreement. The views and opinions expressed herein do not necessarily reflect those of the European Commission. This work was also part-funded by the RCUK Energy Programme under grant EP/I501045.

* See the Appendix of F. Romanelli et al., Proceedings of the 23rd IAEA Fusion Energy Conference 2010, Daejeon, Korea## References

Abdallah J, Clark R E H, Peek J M and Fontes C J, 1994, J. Quant. Spectrosc. Radiat. Transf., **51**, 1
Aggarwal K M and Keenan F P, 2004, Phys. Scr., **69**, 385
Aggarwal K M, Kato T, Keenan F P and Murakami I, 2011, Phys. Scr., **83**, 015302
Badnell N R and Griffin D C, 2000, J. Phys. B, **33** 295524

Bell K L, Gilbody H B, Hughes J G, Kingston A E and Smith F J, 1983, J. Phys. Chem. Ref. Data, **12**, 891
Berrington K A, 1985, J. Phys. B, **18**, L395
Berrington K A, Burke V M, Burke P G and Scialla S, 1989, J. Phys. B, **22**, 665
Berrington K A, Pelan J and Quigley L, 1997, J. Phys. B, **30**, 4973
Brezinsek S *et al.*, 2012, Proc. of the 20th Int. Conf. on Plasma Surface Interactions, May 2012, Aachen, Germany
Crandall D H, Phaneuf R A, Hasselquist B E and Gregory D C, 1979, J. Phys. B, **12**, L249
Dere K P, 2007, Astron. and Astrophys., **466**, 771
Fang D, Hu W, Chen C, Wang Y, Lu F, Tang J and Yang F, 1995, ADNDT, **61**, 91
Fenstermacher M E *et al.*, 1997, Phys. Plasmas, **4**, 1761
Fogle M *et al.*, 2008, Astrophys. J. Suppl. Ser., **175**, 543
Fonck R J, Ramsey A T, Yelle R V, 1982, Appl. Opt., **21**, 2115
Giroud C *et al.*, 2007, Nucl. Fusion, **47**, 313
Golden L B and Sampson D H, 1977, J. Phys. B, **10**, 2229
Golden L B and Sampson D H, 1980, J. Phys. B, **13**, 2645
Griffin D C *et al.*, 2005, J. Phys. B, **38**, L199
Gu M F, 2003, Astrophys. J., **582**, 1241
Kim Y and Rudd M E, 1994, Phys. Rev. A, **50**, 3954
Knopp H, Teng H, Ricz S, Schippers S and Müller A, 2001, Phys. Scr., **T92**, 379
Lawson K D, Zacks J, Coffey I H, Aggarwal K M, Keenan F P and JET-EFDA contributors, 2008, JET Report, JET-RE(08)14, 'Comparison of modelled C VUV line intensity ratios with observations of the emission from the JET plasma SOL - I'
Lawson K D, Zacks J, Coffey I H and JET-EFDA contributors, 2009, JET Report, JET-RE(09)5, 'A relative sensitivity calibration of the JET KT7/2 spectrometer'
Lawson K D, Aggarwal K M, Coffey I H, Keenan F P, O'Mullane M G, Ryć L and Zacks J, 2011, Plasma Phys. Control. Fusion, **53,** 015002
Lawson K D, Aggarwal K M, Coffey I H, Keenan F P, Reid R H G and Zacks J, 2012, Proc. of the 17th Int. Conf. on Atomic Processes in Plasmas, July 2011, Belfast, UK, AIP Conf. Proc. **1438**, 175
Lee T-G, Loch S D, Ballance C P, Ludlow J A and Pindzola M S, 2010, Phys. Rev. A, **82**, 042721
Linkemann J, Müller A, Kenntner J, Habs D, Schwalm D, Wolf A, Badnell N R and Pindzola M S, 1995, Phys. Rev. Lett., **74**, 4173
Loarte A *et al.*, 2007, Nucl. Fusion, **47**, S203
Loch S D *et al.*, 2004, Phys. Rev. E, **69** 066405
Maggi C F, 1996, Ph.D. Thesis, University of Strathclyde
Matthews G F *et al.*, 2011, Phys. Scr., **T145**, 014001
Mitnik D M, Pindzola M S, Griffin D C and Badnell N R, 1999, J. Phys. B, **32** L479
Mitnik D M, Griffin D C, Ballance C P and Badnell N R, 2003, J. Phys. B, **36** 717
Nakano T *et al.*, 2007, Nucl. Fusion, **47**, 1458
Nakano T *et al.*, 2009, J. Nucl. Materials, **390-391**, 255
Pindzola M S, Ballance C P and Loch S D, 2011, Phys. Rev. A, **83**, 062705
Ralchenko Yu, Kramida A E, Reader J and NIST ASD Team, 2011, NIST Atomic Spectra Database, ver. 4.1.0, 'http://physics.nist.gov/asd3'
Ryans R S I, Foster-Woods V J, Copeland F, Keenan F P and Matthews A, 1998, ADNDT, **70**, 179
Sampson D H and Zhang H L, 1988, Phys. Rev. A, **37**, 3765
Schwob J L, Wouters A W, Suckewer S and Finkenthal M, 1987, Rev. Sci. Instrum., **58**, 1601
Scott M P, Teng H and Burke P G, 2000, J. Phys. B, **33** L63
Summers H P, 2004, 'The ADAS User Manual, version 2.6, 'http://adas.phys.strath.ac.uk'
Suno H and Kato T, 2005, NIFS-DATA-91
Wolf R C *et al.*, 1995, JET Preprint, JET-P(95)34
Younger S M, 1981, Phys. Rev. A, **24**, 1278

**Figure captions**

Figure 1. Schematic diagram of energy levels in adjacent ionization stages, showing the main populating channels of level 2 in the higher ionization stage, with: ⎯ electron collision excitation; ⋯⋯ radiative decay; -··- recombination; - - - ionization.

Figure 2. Ionization PECs for C IV transitions plotted as a function of $T_e$, with: ⎯ for the 244.9 Å transition; ⋯⋯ for 289.2 Å; - - - for 296.9 Å; -··- for 312.4 Å; -··- for 384.1 Å; -×- for 419.6 Å; ⎯⎯ for 1168.9 Å; -*- for 1548.2 Å; -+- for 1550.8 Å.

Figure 3. Ratios of C IV ionization/excitation PECs for different values of $n_{g-1}/n_g$ plotted as a function of $T_e$, with: -··- for the 296.9 Å transition; -··- for 312.4 Å and 5805 Å; - - - for 419.6 Å; ⋯⋯ for 1548.2 Å; ⎯ for 1550.8 Å. Note that for no symbol $n_{g-1}/n_g = 1$, while + is for $n_{g-1}/n_g = 3$ and × is for $n_{g-1}/n_g = 10$.

Figure 4. Ionization PECs for C V transitions plotted as a function of $T_e$, with: ⎯ for the 189.3 Å transition; ⎯⎯ for 198.1 Å; ⋯⋯ for 260.2 Å; - - - for 271.9 Å; -··- for 756.9 Å; -··- for 775.9 Å.

Figure 5. Ratios of C V ionization/excitation PECs for different values of $n_{g-1}/n_g$ plotted as a function of $T_e$, with: ⎯ for the 189.3 Å and 756.9 Å transitions; ⎯⎯ for 198.1 Å and 775.9 Å; -··- for 260.2 Å; - - - for 271.9 Å. Note that for no symbol $n_{g-1}/n_g = 0.2$, while + is for $n_{g-1}/n_g = 1$ and × is for $n_{g-1}/n_g = 3$.

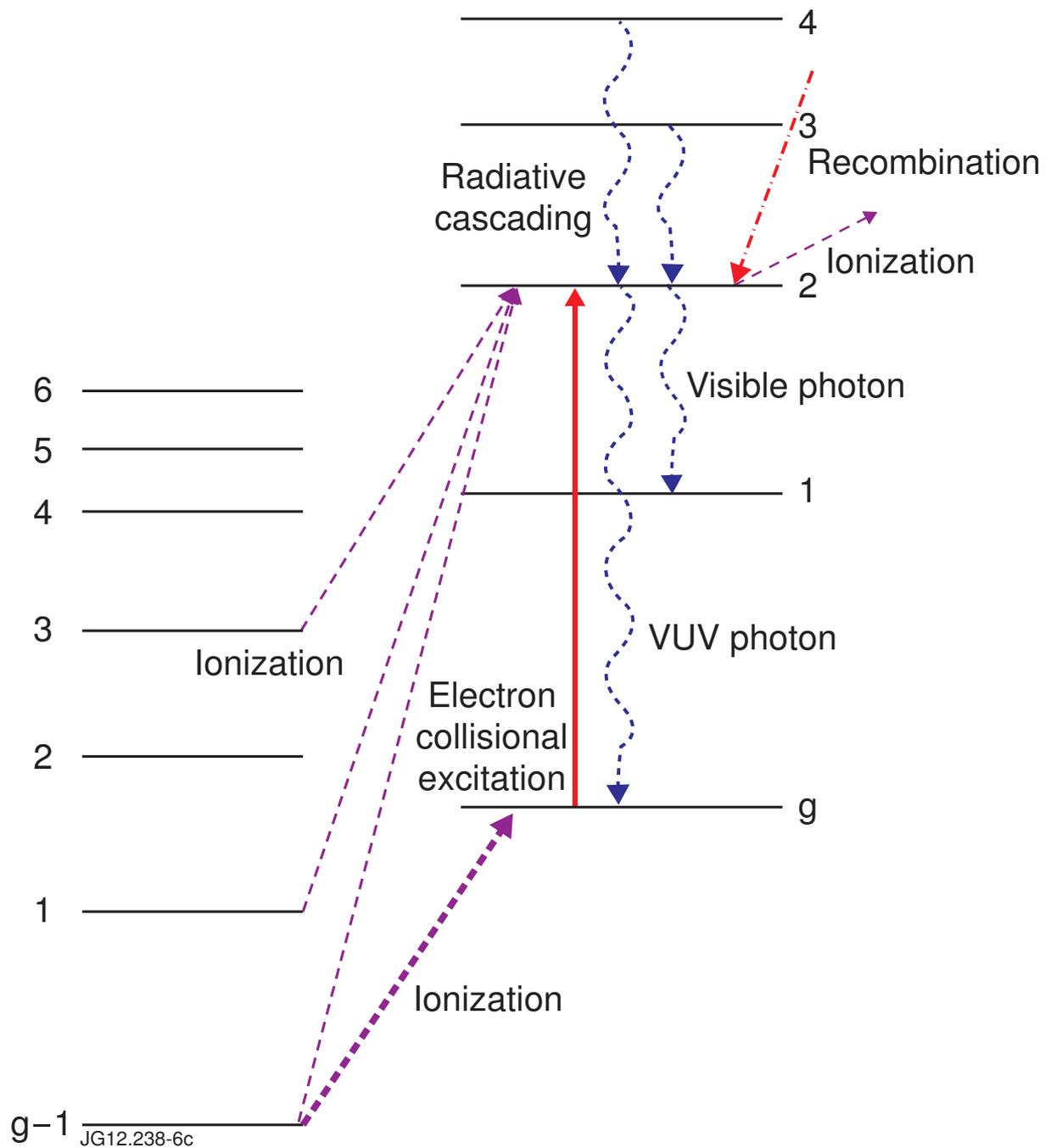

Figure 1.

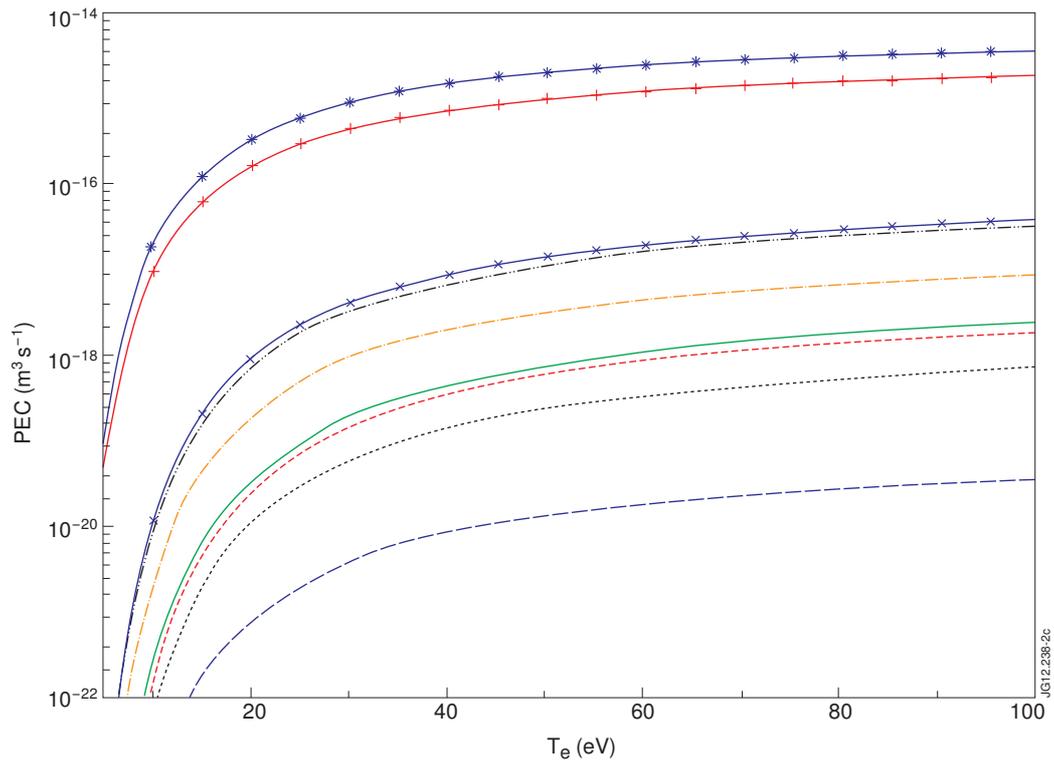

Figure 2.

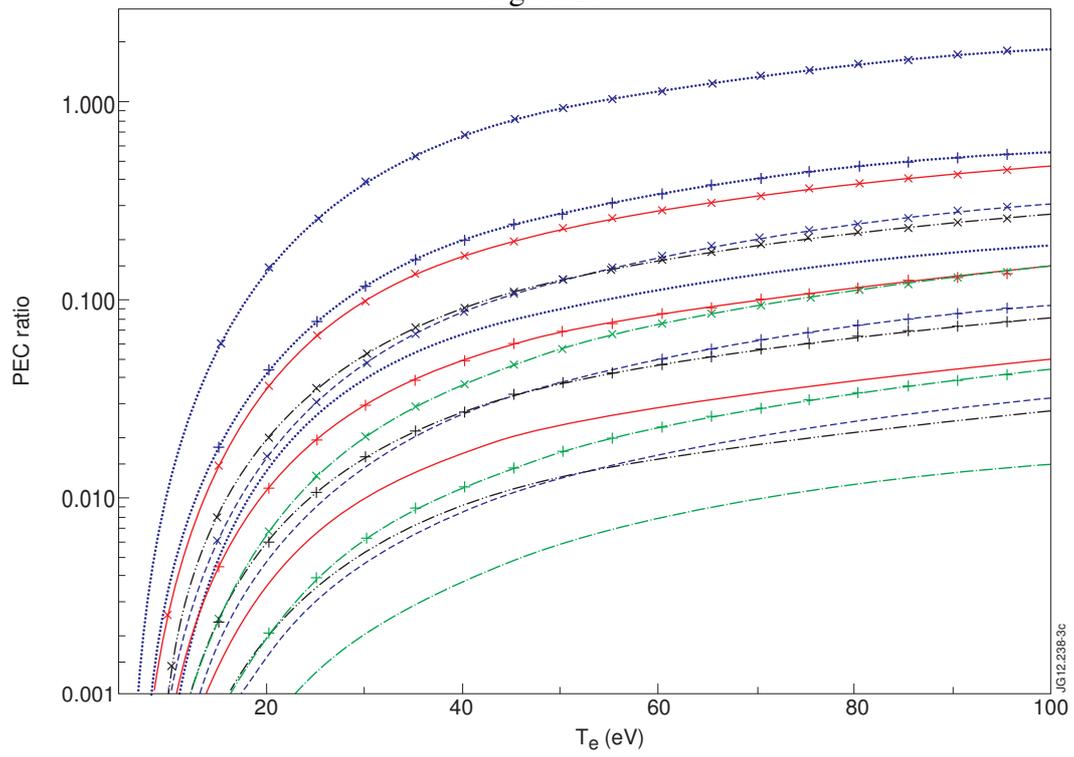

Figure 3.

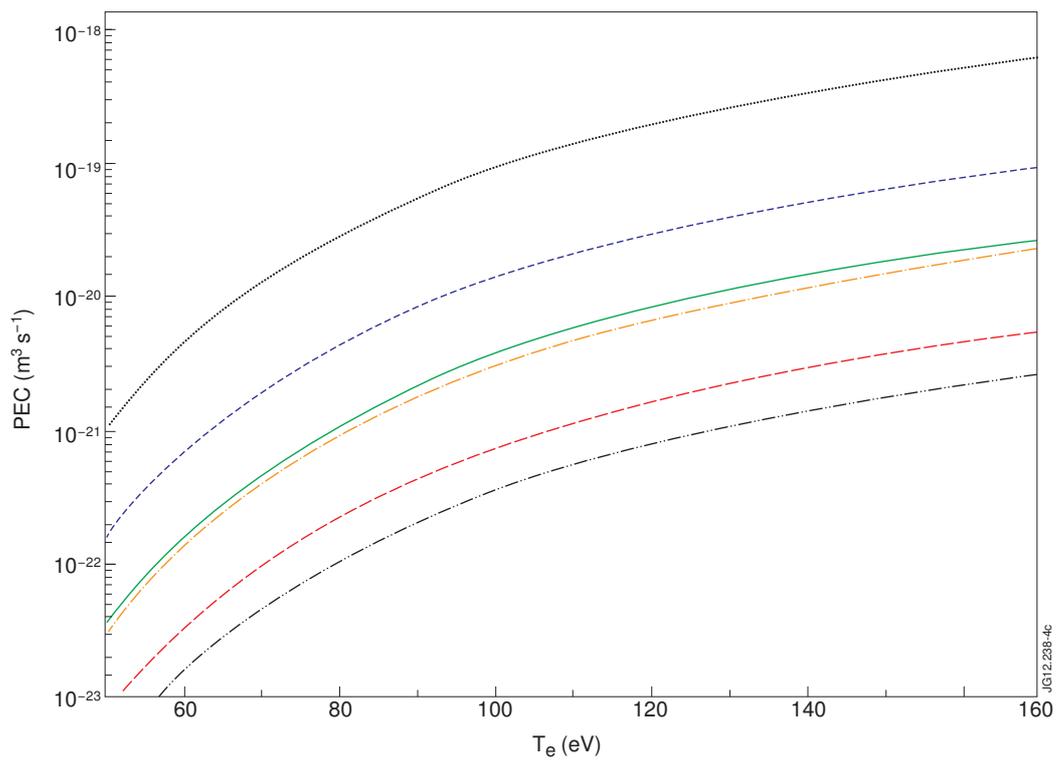

Figure 4.

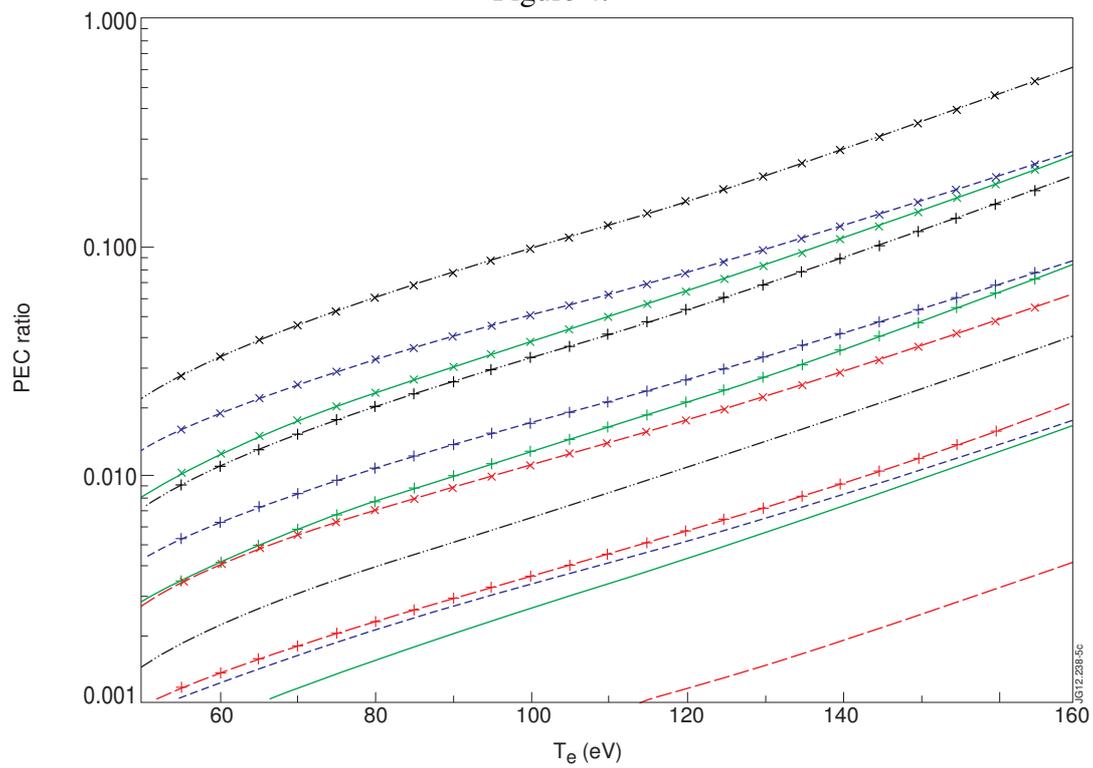

Figure 5.